%% file: main.tex
\documentclass[superscriptaddress,twocolumn,nopreprintnumbers,floatfix,nofootinbib,trackchanges,table]{aastex631}
\pdfoutput=1

\usepackage{amsmath}

\usepackage[T1]{fontenc} 
\usepackage[utf8]{inputenc}
\usepackage{amssymb,graphicx}
\usepackage{scrextend}
\usepackage[caption=false]{subfig}

\usepackage{multirow}

\usepackage{siunitx}
\DeclareSIUnit\parsec{pc}
\DeclareSIUnit\Mpc{\mega\parsec}
\DeclareSIUnit\yr{yr}

\usepackage{txfonts}
\newlength{\capheight}
\usepackage{textgreek}

\usepackage{xspace}
\usepackage{array}
\usepackage{booktabs}
\usepackage{color}

\usepackage{acronym}
\input{acronyms}

\input{units_symbols}

\input{results_numbers_macros}

\definecolor{danger-red}{rgb}{0.8, 0.4, 0.0}
\definecolor{warning-amber}{rgb}{0.9, 0.6, 0.0}
\definecolor{ok-green}{rgb}{0.0, 0.6, 0.5}
\newcommand{\reviewed}[1]{{#1}}

\begin{document}

\title{Searches for Binary Mergers with Sub-solar Mass Components in Data from the First Part of LIGO--Virgo--KAGRA's Fourth Observing Run}

\input{authors_LVK_Aug2025_aas_v2.tex}

\date[\relax]{Compiled: \today}

\begin{abstract}
\input{abstract}
\end{abstract}


\input{introduction}
\input{search}

\input{search_sensitivity}
\input{rates}
\input{dm_constraints}
\input{conclusion}

\section*{Data availability}
\input{data.tex}

\section*{Acknowledgements}
\input{LVKack}
\facility{LIGO}
\software{\input{software_ack.tex}}

\appendix
\input{VT_methods}
\clearpage
\bibliographystyle{aasjournal}
\bibliography{references.bib}

\end{document}

%% file: acronyms.tex
\acrodef{LSC}[LSC]{LIGO Scientific Collaboration}
\acrodef{LVC}[LVC]{LIGO Scientific and Virgo Collaboration}
\acrodef{LVK}[LVK]{LIGO--Virgo--KAGRA Collaboration}
\acrodef{aLIGO}{Advanced Laser Interferometer Gravitational-Wave Observatory}
\acrodef{AdV}{Advanced Virgo}
\acrodef{LIGO}[LIGO]{Laser Interferometer Gravitational-Wave Observatory}
\acrodef{IFO}[IFO]{interferometer}
\acrodef{LHO}[LHO]{LIGO Hanford Observatory}
\acrodef{LLO}[LLO]{LIGO Livingston Observatory}
\acrodef{O2}[O2]{second observing run}
\acrodef{O1}[O1]{first observing run}
\acrodef{O3}[O3]{third observing run}
\acrodef{O3a}[O3a]{first half of the third observing run}
\acrodef{O3b}[O3b]{second half of the third observing run}
\acrodef{O4}[O4]{fourth observing run}
\acrodef{O4a}[O4a]{first part of the fourth observing run}
\acrodef{O4b}[O4b]{second part of the fourth observing run}
\acrodef{O4c}[O4c]{third part of the fourth observing run}
\acrodef{GWTC}[GWTC]{Gravitational Wave Transient Catalog}
\acrodef{GWOSC}[GWOSC]{Gravitational Wave Open Science Center}

\acrodef{BH}[BH]{black hole}
\acrodef{BBH}[BBH]{binary black hole}
\acrodef{BNS}[BNS]{binary neutron star}
\acrodef{IMBH}[IMBH]{intermediate-mass black hole}
\acrodef{NS}[NS]{neutron star}
\acrodef{BHNS}[BHNS]{black hole--neutron star binaries}
\acrodef{NSBH}[NSBHs]{neutron star--black hole binaries}
\acrodef{PBH}[PBH]{primordial black hole}
\acrodef{DBH}[DBH]{dark black hole}
\acrodef{CBC}[CBC]{compact binary coalescence}
\acrodef{GW}[GW]{gravitational wave}
\acrodef{GWH}[GW]{gravitational-wave}
\acrodef{SSM}[SSM]{sub-solar mass}

\acrodef{CWB}[cWB]{coherent WaveBurst}

\acrodef{SNR}[SNR]{signal-to-noise ratio}
\acrodef{FAR}[FAR]{false alarm rate}
\acrodef{IFAR}[IFAR]{inverse false alarm rate}
\acrodef{FAP}[FAP]{false alarm probability}
\acrodef{PSD}[PSD]{noise power spectral density}
\acrodef{SVD}[SVD]{singular value decomposition}
\acrodef{GR}[GR]{general relativity}
\acrodef{NR}[NR]{numerical relativity}
\acrodef{PN}[PN]{post-Newtonian}
\acrodef{EOB}[EOB]{effective-one-body}
\acrodef{ROM}[ROM]{reduced-order model}
\acrodef{IMR}[IMR]{inspiral--merger--ringdown}

\acrodef{PDF}[PDF]{probability density function}
\acrodef{PE}[PE]{parameter estimation}
\acrodef{CL}[CL]{credible level}

\acrodef{EOS}[EoS]{equation of state}

\acrodef{LAL}[LAL]{LIGO Algorithm Library}

\acrodef{KLD}[KLD]{Kullback--Leibler divergence}
\acrodef{JSD}[JSD]{Jensen--Shannon divergence}

\newcommand{\GSTLAL}{\textsc{GstLAL}\xspace}
\newcommand{\MANIFOLD}{\textsc{manifold}\xspace}

\newcommand{\PYCBC}{\textsc{PyCBC}\xspace}
\newcommand{\MBTA}{\textsc{MBTA}\xspace}

\newcommand{\LALSUITE}{\textsc{LALSuite}\xspace}

\newcommand{\GWPOPULATION}{\textsc{GWPopulation}\xspace}

\newcommand{\NUMPY}{\textsc{NumPy}\xspace}
\newcommand{\SCIPY}{\textsc{SciPy}\xspace}

\newcommand{\SEABORN}{\textsc{seaborn}\xspace}
\newcommand{\GWPY}{\textsc{GWpy}\xspace}

\newcommand{\MATPLOTLIB}{\textsc{Matplotlib}\xspace}
\newcommand{\PANDAS}{\textsc{Pandas}\xspace}

%% file: units_symbols.tex
\newcommand{\Mc}{\ensuremath{\mathcal{M}}\xspace}

\newcommand{\Msun}{\ensuremath{M_\odot}\xspace}
\newcommand{\massone}{\ensuremath{m_1}\xspace}
\newcommand{\masstwo}{\ensuremath{m_2}\xspace}

\newcommand{\masscomponent}{\ensuremath{m_{1,2}}\xspace}

\newcommand{\spinone}{\ensuremath{\chi_1}\xspace}

\newcommand{\spintwo}{\ensuremath{\chi_2}\xspace}

\newcommand{\spincomponent}{\ensuremath{\chi_{1,2}}\xspace}

\newcommand{\redshift}{\ensuremath{z}\xspace}



%% file: results_numbers_macros.tex
\newcommand{\RUNSTART}{\reviewed{24 May 2023 15:00:00 UTC}\xspace}      
\newcommand{\RUNEND}{\reviewed{16 January 2024 16:00:00 UTC}\xspace}    
\newcommand{\ERSTART}{\reviewed{15 May 2023 14:13:22 UTC}\xspace}       
\newcommand{\EREND}{\reviewed{19 May 2023 17:31:33 UTC}\xspace}         

\newcommand{\RateULMin}{\reviewed{\ensuremath{110\,{\rm Gpc^{-3}\,yr^{-1}}}}} 
\newcommand{\RateULMax}{\reviewed{\ensuremath{10000\,{\rm Gpc^{-3}\,yr^{-1}}}}} 
\newcommand{\BNSVTorderofmagnitude}{\reviewed{\qty{\sim e-3}{\giga\parsec\cubed\yr}}} 
\newcommand{\LBfPBHBelowOneOfoura}{\reviewed{0.9\,\Msun}}
\newcommand{\EBfPBHMaxMPBHpercentOfoura}{\reviewed{7\%}}
\newcommand{\EBfPBHMinMPBHpercentOfoura}{\reviewed{40\%}}

\newcommand{\SignificantOPA}{\reviewed{\ensuremath{2~\mathrm{yr}^{-1}}}\xspace}

\newcommand{\MOneMIN}{\reviewed{0.2\,\Msun}\xspace}
\newcommand{\MOneMAX}{\reviewed{10.0\,\Msun}\xspace}
\newcommand{\MTwoMIN}{\reviewed{0.2\,\Msun}\xspace}
\newcommand{\MTwoMAX}{\reviewed{1.0\,\Msun}\xspace}
\newcommand{\MassRatioMin}{\reviewed{0.1}\xspace}
\newcommand{\MassRatioMax}{\reviewed{1.0}\xspace}
\newcommand{\MaxSpinLowMass}{\reviewed{0.1}\xspace}
\newcommand{\MaxSpinHighMass}{\reviewed{0.9}\xspace}
\newcommand{\SSMLowHighMassGap}{\reviewed{0.5}\xspace} 
\newcommand{\MinMatch}{\reviewed{0.97}\xspace}

\newcommand{\IMRPHENOMD}{\reviewed{IMRPhenomD}\xspace}

\newcommand{\TAYLORFTWO}{\reviewed{TaylorF2}\xspace} 
\newcommand{\SEOBNRVFOURROM}{\reviewed{SEOBNRv4ROM}\xspace}

\newcommand{\GSTLALSSMMinMatch}{\reviewed{0.965}\xspace}
\newcommand{\GstlalFLOW}{\reviewed{\ensuremath{45\,\mathrm{Hz}}}\xspace}
\newcommand{\GstlalFHIGH}{\reviewed{\ensuremath{1024\,\mathrm{Hz}}}\xspace}
\newcommand{\GstlalMAXDUR}{\reviewed{\ensuremath{128\,\mathrm{s}}}\xspace}

\newcommand{\GstlalDtDphiFHIGH}{\reviewed{\ensuremath{512\,\mathrm{Hz}}}\xspace}

\newcommand{\MBTASEARCHAPPROX}{\reviewed{SpinTaylorT4}\xspace}
\newcommand{\MBTAFLOW}{\reviewed{\ensuremath{45\,\mathrm{Hz}}}\xspace}
\newcommand{\MBTAFHIGH}{\reviewed{\ensuremath{1000\,\mathrm{Hz}}}\xspace}

\newcommand{\PyCBCBankFLOW}{\reviewed{\ensuremath{45\,\mathrm{Hz}}}\xspace}
\newcommand{\PyCBCBankFHIGH}{\reviewed{\ensuremath{1000\,\mathrm{Hz}}}\xspace}
\newcommand{\PyCBCFilterFLOW}{\reviewed{\ensuremath{30\,\mathrm{Hz}}}\xspace}
\newcommand{\PyCBCMaxTempltDur}{\reviewed{\ensuremath{1008\,\mathrm{s}}}\xspace}
\newcommand{\PyCBCMtotThr}{\reviewed{1.0\,\Msun}\xspace}

\newcommand{\GSTLALSecondTriggerTime}{\reviewed{2023 May 28 11:07:32.09 UTC}\xspace}

\newcommand{\GSTLALTableRow}[8]{%
  \reviewed{\ensuremath{#1}} & 
  \GSTLAL &                    
  \reviewed{#2} &              
  \reviewed{\ensuremath{#3}} & 
  \reviewed{\ensuremath{#4}} & 
  \reviewed{\ensuremath{#5}} & 
  \reviewed{\ensuremath{#6}} & 
  \reviewed{#7} &              
  \reviewed{\ensuremath{#8}} \\
}

\newcommand{\GSTLALTrigOneData}{%
  {0.00052}         
  {2023 May 29 18:15:00.75 UTC} 
  {6.49}            
  {0.98}            
  {0.27}            
  {0.27}            
  {L}               
  {10.63}           
}

\newcommand{\GSTLALTrigTwoData}{%
  {1.2}            
  {2023 May 28 11:07:32.09 UTC} 
  {6.49}            
  {0.98}            
  {0.73}            
  {0.73}            
  {HL}              
  {9.48}            
}

\newcommand{\MBTATableRow}[8]{%
  \reviewed{\ensuremath{#1}} & 
  \MBTA &                      
  \reviewed{#2} &              
  \reviewed{\ensuremath{#3}} & 
  \reviewed{\ensuremath{#4}} & 
  \reviewed{\ensuremath{#5}} & 
  \reviewed{\ensuremath{#6}} & 
  \reviewed{#7} &              
  \reviewed{\ensuremath{#8}} \\
}

\newcommand{\MBTATrigOneData}{%
  {0.87}            
  {2023 September 14 21:02:19.50 UTC} 
  {2.63}            
  {0.29}            
  {-0.72}           
  {-0.02}           
  {HL}              
  {9.44}            
}

\newcommand{\MBTATrigTwoData}{%
  {0.91}            
  {2023 May 29 18:15:00.75 UTC} 
  {7.64}            
  {0.88}            
  {0.39}            
  {-0.07}           
  {L}               
  {9.96}            
}

\newcommand{\MBTATrigThreeData}{%
  {1.3}            
  {2023 October 14 08:15:06.33 UTC} 
  {2.28}            
  {0.22}            
  {0.90}            
  {0.05}            
  {HL}              
  {10.28}           
}

\newcommand{\MBTATrigFourData}{%
  {1.3}            
  {2023 August 10 07:13:47.88 UTC} 
  {1.86}            
  {0.32}            
  {0.65}            
  {-0.10}           
  {L}               
  {12.16}           
}

\newcommand{\PYCBCTableRow}[8]{%
  \reviewed{\ensuremath{#1}} & 
  \PYCBC &                        
  \reviewed{#2} &               
  \reviewed{\ensuremath{#3}} & 
  \reviewed{\ensuremath{#4}} & 
  \reviewed{\ensuremath{#5}} & 
  \reviewed{\ensuremath{#6}} & 
  \reviewed{#7} &              
  \reviewed{\ensuremath{#8}} \\
}

\newcommand{\PYCBCTrigOneData}{%
  {0.0010}           
  {2023 May 29 18:15:00.75 UTC} 
  {7.73}            
  {0.87}            
  {0.43}            
  {-0.64}           
  {L}               
  {11.10}           
}

\newcommand{\PYCBCTrigTwoData}{%
  {1.3}            
  {2023 July 19 11:50:50.27 UTC} 
  {0.74}            
  {0.24}            
  {-0.24}           
  {0.10}            
  {HL}              
  {9.47}            
}

\newcommand{\PYCBCTrigThreeData}{%
  {1.4}            
  {2023 August 10 10:10:03.37 UTC} 
  {0.60}            
  {0.21}            
  {0.87}            
  {-0.07}           
  {HL}              
  {9.43}            
}

\newcommand{\GSTLALLoudestTriggerFAR}{\reviewed{\ensuremath{1.2\,\mathrm{yr^{-1}}}}} 
\newcommand{\MBTALoudestTriggerFAR}{\reviewed{\ensuremath{0.87\,\mathrm{yr^{-1}}}}} 
\newcommand{\PYCBCLoudestTriggerFAR}{\reviewed{\ensuremath{1.3\,\mathrm{yr^{-1}}}}} 

\newcommand{\FourthAObsAnalysisTime}{\reviewed{\ensuremath{0.76\,\mathrm{yr}}}} 
\newcommand{\ThirdObsAnalysisTime}{\reviewed{\ensuremath{0.9\,\mathrm{yr}}}} 

\newcommand{\TotalInjVT}{\reviewed{\ensuremath{1.65\,\mathrm{Gpc^{3}\,yr}}}} 

\newcommand{\InjSpacing}{\reviewed{\ensuremath{32}\xspace}} 
\newcommand{\MBTAInjSpacing}{\reviewed{\ensuremath{16}\xspace}} 

\newcommand{\InjMassPowerLawAlpha}{\reviewed{\ensuremath{-1}}}
\newcommand{\MinPrimaryMassBH}{\reviewed{\ensuremath{0.19\,\Msun}}}
\newcommand{\MaxPrimaryMassBH}{\reviewed{\ensuremath{11.1\,\Msun}}}
\newcommand{\MinSecondaryMassBH}{\reviewed{\ensuremath{0.19\,\Msun}}}
\newcommand{\MaxSecondaryMassBH}{\reviewed{\ensuremath{1.1\,\Msun}}}
\newcommand{\MinPrimaryMassNS}{\reviewed{\ensuremath{0.5\,\Msun}}}
\newcommand{\MaxPrimaryMassNS}{\reviewed{\ensuremath{3\,\Msun}}}
\newcommand{\MinSecondaryMassNS}{\reviewed{\ensuremath{0.5\,\Msun}}}
\newcommand{\MaxSecondaryMassNS}{\reviewed{\ensuremath{1.1\,\Msun}}}

\newcommand{\DistributionWeight}{\reviewed{\ensuremath{0.5}}} 
\newcommand{\NormalDistMu}{\reviewed{\ensuremath{0}}}
\newcommand{\NormalDistSigma}{\reviewed{\ensuremath{0.1}}}
\newcommand{\BetaDistAlpha}{\reviewed{\ensuremath{1}}}
\newcommand{\BetaDistBeta}{\reviewed{\ensuremath{4}}}
\newcommand{\MinSpinMag}{\reviewed{\ensuremath{0}}}
\newcommand{\MaxSpinMag}{\reviewed{\ensuremath{1}}}

\newcommand{\IMRPHENOMXPHM}{\reviewed{\textsc{IMRPhenomXPHM}}\xspace} 
\newcommand{\IMRPHENOMXPNRTIDALVTHREE}{\reviewed{\textsc{IMRPhenomXP\_NRTidalv3}}\xspace} 
\newcommand{\GPPVADDtwo}{\reviewed{\textsc{GPPVA+DD2}}\xspace} 

\newcommand{\MaxRedshiftInj}{\reviewed{\ensuremath{0.2}\xspace}} 

\newcommand{\MaxSpinMagVTpop}{\reviewed{\ensuremath{0.1}}} 
\newcommand{\GstLALVTMcratio}{\reviewed{\ensuremath{1.1}}}
\newcommand{\MBTAVTMcratio}{\reviewed{\ensuremath{1.0}}}
\newcommand{\PyCBCVTMcratio}{\reviewed{\ensuremath{1.3}}}
\newcommand{\GstLALVTCompMassratio}{\reviewed{\ensuremath{1.2}}}
\newcommand{\MBTAVTCompMassratio}{\reviewed{\ensuremath{1.0}}}
\newcommand{\PyCBCVTCompMassratio}{\reviewed{\ensuremath{1.4}}}

\newcommand{\LogNormSigmaBNSVT}{\reviewed{\ensuremath{0.1}}} 
\newcommand{\maxspinBNS}{\reviewed{\ensuremath{0.4}}} 

\newcommand{\BNSVTFARThreshold}{\reviewed{\ensuremath{1~{\rm yr^{-1}}}}} 

\newcommand{\MostConstrainedRateMcGSTLAL}{\reviewed{\ensuremath{130\,{\rm Gpc^{-3}\,yr^{-1}}}}}
\newcommand{\MostConstrainedRateMcMBTA}{\reviewed{\ensuremath{140\,{\rm Gpc^{-3}\,yr^{-1}}}}}
\newcommand{\MostConstrainedRateMcPYCBC}{\reviewed{\ensuremath{120\,{\rm Gpc^{-3}\,yr^{-1}}}}}
\newcommand{\MostConstrainedRateCompmassGSTLAL}{\reviewed{\ensuremath{130\,{\rm Gpc^{-3}\,yr^{-1}}}}}
\newcommand{\MostConstrainedRateCompmassMBTA}{\reviewed{\ensuremath{130\,{\rm Gpc^{-3}\,yr^{-1}}}}}
\newcommand{\MostConstrainedRateCompmassPYCBC}{\reviewed{\ensuremath{110\,{\rm Gpc^{-3}\,yr^{-1}}}}}

\newcommand{\BNSRateULMin}{\reviewed{\ensuremath{570}}}
\newcommand{\BNSRateULMax}{\reviewed{\ensuremath{710}}}

\newcommand{\BNSRatePYCBC}{\reviewed{\ensuremath{570}}}
\newcommand{\BNSRateMBTA}{\reviewed{\ensuremath{710}}}
\newcommand{\BNSRateGSTLAL}{\reviewed{\ensuremath{610}}}

\newcommand{\MinMPBH}{\reviewed{\ensuremath{0.35\,\Msun}}}
\newcommand{\EBfPBHMinMPBHOfoura}{\reviewed{\ensuremath{0.4}}}
\newcommand{\MaxMPBH}{\reviewed{\ensuremath{1\,\Msun}}}
\newcommand{\EBfPBHMaxMPBHOfoura}{\reviewed{\ensuremath{0.07}}}
\newcommand{\LBmPBHfPBHconstrainedOfoura}{\reviewed{0.9\,\Msun}}

\newcommand{\fconstraintOfoura}{\reviewed{(0.9\,\mathrm{to}\,1.2) \times 10^{-5}}}
\newcommand{\darkprotonlower}{\reviewed{2.4}}
\newcommand{\darkprotonupper}{\reviewed{\ensuremath{16.5}\,\text{--}\,\ensuremath{16.6}}}
\newcommand{\mminlowerbound}{\reviewed{\ensuremath{0.020\,\Msun}\,\text{--}\,\ensuremath{0.021\,\Msun}}\xspace}
\newcommand{\FARThreshVTDBH}{\reviewed{\ensuremath{0.01~\mathrm{yr}^{-1}}}}



%% file: authors_LVK_Aug2025_aas_v2.tex


\author[0000-0003-4786-2698]{A.~G.~Abac}
\affiliation{Max Planck Institute for Gravitational Physics (Albert Einstein Institute), D-14476 Potsdam, Germany}
\author{I.~Abouelfettouh}
\affiliation{LIGO Hanford Observatory, Richland, WA 99352, USA}
\author{F.~Acernese}
\affiliation{Dipartimento di Farmacia, Universit\`a di Salerno, I-84084 Fisciano, Salerno, Italy}
\affiliation{INFN, Sezione di Napoli, I-80126 Napoli, Italy}
\author[0000-0002-8648-0767]{K.~Ackley}
\affiliation{University of Warwick, Coventry CV4 7AL, United Kingdom}
\author{A.~Adam}
\affiliation{OzGrav, University of Western Australia, Crawley, Western Australia 6009, Australia}
\author[0000-0001-5525-6255]{C.~Adamcewicz}
\affiliation{OzGrav, School of Physics \& Astronomy, Monash University, Clayton 3800, Victoria, Australia}
\author[0009-0004-2101-5428]{S.~Adhicary}
\affiliation{The Pennsylvania State University, University Park, PA 16802, USA}
\author{D.~Adhikari}
\affiliation{Max Planck Institute for Gravitational Physics (Albert Einstein Institute), D-30167 Hannover, Germany}
\affiliation{Leibniz Universit\"{a}t Hannover, D-30167 Hannover, Germany}
\author[0000-0002-4559-8427]{N.~Adhikari}
\affiliation{University of Wisconsin-Milwaukee, Milwaukee, WI 53201, USA}
\author[0000-0002-5731-5076]{R.~X.~Adhikari}
\affiliation{LIGO Laboratory, California Institute of Technology, Pasadena, CA 91125, USA}
% ...

\collaboration{3000}{The LIGO Scientific Collaboration, the Virgo Collaboration, and the KAGRA Collaboration}



%% file: abstract.tex
We report on a gravitational wave search for compact binary coalescences involving at least one component with mass between $0.2\,\Msun$ to $1\,\Msun$, and ratio of component masses between 0.1 and 1.
The analysis uses data collected by the LIGO detectors between May 24 2023 15:00 UTC and January 16 2024 16:00 UTC.
No statistically significant sub-solar mass candidates were identified by the participating search algorithms. 
We report the detection sensitivity of the current searches to the target sub-solar mass black hole population.
With the absence of detections, we place upper limits on the merger rate of sub-solar mass black holes, ranging from \RateULMin~to~\RateULMax~at 90\% confidence. 
We constrain two illustrative dark matter scenarios that can form sub-solar mass compact objects with these searches: primordial black holes, and dark black holes forming in a dissipative dark matter model.
For late-forming primordial black hole binaries, our search excludes the fraction of dark matter in primordial black holes to be $\leq 1$ only for masses above $\LBfPBHBelowOneOfoura$.
In the early-formation scenario, we limit this fraction to be $\leq\EBfPBHMaxMPBHpercentOfoura$ at $\MaxMPBH$, and $\leq\EBfPBHMinMPBHpercentOfoura$ at $\MinMPBH$.
For the dissipative model, the excluded region in the parameter space of dark matter fraction in dark black holes and their minimum possible mass extends down to $\fconstraintOfoura$ at $1\,\Msun$ with no constraints below $0.02\,\Msun$.
For the first time, we report the detection sensitivity of our searches to binaries with sub-solar mass neutron stars, and place the 90\% confidence merger rate limit at (\BNSRateULMin~to~\BNSRateULMax)~${\rm Gpc^{-3}\,yr^{-1}}$ for a population with component masses distributed uniformly down to $0.5\,\Msun$.

%% file: introduction.tex
\section{Introduction}\label{sec:intro}

The \acl{LVK}~\citep[\acsu{LVK};][]{KAGRA:2013rdx} has published over 300 probable \ac{GW} candidates from searches for \acp{CBC} in data from the \acl{O1}~\citep[\acsu{O1};][]{LIGOScientific:2018mvr}, \acl{O2}~\citep[\acsu{O2};][]{LIGOScientific:2020ibl,LIGOScientific:2021usb}, \acl{O3}~\citep[\acs{O3};][]{KAGRA:2021vkt}, the \acl{O4a}~\citep[\acsu{O4a};][]{LIGOScientific:2025slb,LIGOScientific:2025snk}, and the \acl{O4b}~\citep[\acsu{O4b};][]{LIGOScientific:2026jgl,LIGOScientific:2026wfs} of the \acl{LIGO}~\citep[\acsu{LIGO};][]{PhysRevD.111.062002,PhysRevX.13.041021,Jia:2024iqe,Soni2025,PhysRevD.93.112004,Harry2010,PhysRevLett.116.131103} and Virgo~\citep{Acernese2015} detectors in the latest Gravitational-Wave Transient Catalog, GWTC-4.0~\citep{LIGOScientific:2025slb,LIGOScientific:2025hdt,LIGOScientific:2025yae}.
These candidates are produced by searches targeting \acp{BBH}, \acp{BNS} and \ac{NSBH} with component masses $\geq 1\,\Msun$, where the lower bound is motivated observationally by known pulsars, and theoretically by the expectation that neutron star masses should lie near the Chandrasekhar limit for white dwarfs of $1.4\,\Msun$ dictated by the proton mass and neutron mass~\citep{Chandrasekhar:1931ih}.
Additionally, black holes produced by stars are expected to have masses exceeding neutron star masses.

Although \ac{GW} detectors are primarily sensitive to mergers of stellar-mass compact objects, they can also detect signals from less massive binaries of compact objects, if such systems exist in nature~\citep{Magee:2018opb}.
Targeted searches for \ac{SSM} compact objects open a window into exotic formation channels that may involve new physics beyond the standard astrophysical paradigm~\citep{Singh:2024fdd}.
A key motivation for these searches is the possibility that some compact objects could be \acp{PBH} which are a candidate for dark matter~\citep{Hawking:1971ei,Chapline:1975ojl,Green:2020jor}.
\acp{PBH} have been proposed to contribute to the population of \ac{GW} sources since the first detection of GW150914~\citep{LIGOScientific:2016aoc,Bird:2016dcv,Sasaki:2016jop,Clesse:2016vqa}.
They are purported to form in the early universe and dominate the merger rate across all redshifts with a peak around \redshift $\geq 30$, compared to the merger rate of stellar black holes, which, if it follows the star formation rate, would peak around \redshift $\sim$ 1--2~\citep{Ng:2022agi,Ng:2022vbz}.
Even though \acp{PBH} can populate a wide mass spectrum extending above a solar mass~\citep{Carr:2019kxo}, we rely on the detection of a sub-solar mass candidate to conclusively distinguish them from stellar black holes.

\ac{GW} observations of \acp{BBH} can also constrain theories of dissipative dark matter and its chemistry where dark halos gravitationally collapse to form \acp{DBH}~in regions of sufficiently high dark matter density~\citep{Shandera:2018xkn,LIGOScientific:2022hai,Singh:2020wiq}.
The merger rate of \acp{DBH} in this scenario is inversely proportional to the average~\ac{DBH} mass.
Therefore, \ac{SSM} black holes forming from dissipative halos could have a higher predicted merger rate than black holes forming through regular stellar evolutionary channels, which implies that \ac{SSM} searches can provide strict limits on dissipative dark matter models~\citep{Shandera:2018xkn}.
Especially if the lowest possible \ac{DBH} mass is smaller than the white dwarf Chandrasekhar limit, we can start to probe the particle nature of dissipative dark matter using \ac{SSM} searches.

Additionally, models where accumulation of non-annihilating dark matter transmutes \acp{NS} to \acp{BH} with masses below the $2.5\,\Msun$ maximum \ac{NS} mass have been constrained using limits on merger rates from \acsu{O3} \ac{SSM} searches.
Non-detections of these low-mass black holes provide upper limits on the interaction cross-section of dark matter particles with nucleons for GeV scale dark matter~\citep{Dasgupta:2020mqg,Bhattacharya:2023stq}.

Searches for binaries with \ac{SSM} compact objects have been performed in data from LIGO and Virgo since initial LIGO's science runs~\citep{LIGOScientific:2005ad, PhysRevD.77.062002, LIGOScientific:2018glc, LIGOScientific:2019kan, LIGOScientific:2021job, LIGOScientific:2022hai,Nitz:2021mzz, Nitz:2020bdb, Phukon:2021cus, 2022PhRvD.106b3024N, 2025ApJ...984...61K, Kacanja:2026byy}.
\ac{LVK} matched-filtering \ac{SSM} searches have probed masses down to $0.2\,\Msun$; this parameter space initially motivated by the gravitational microlensing surveys of the LMC which suggested that $\sim 20\%$ of the Galactic halo is composed of massive compact halo objects with masses in the range $0.15\,\Msun$ to $0.9\,\Msun$~\citep{LIGOScientific:2005ad}.
Since \ac{O2}, our \ac{SSM} searches have allowed for spinning components, initially with dimensionless spin magnitudes $ \leq 0.1$~\citep{LIGOScientific:2019kan}.
In \ac{O3}, we increased the primary component's mass to 10\,\Msun to include systems with more asymmetric masses, while also increasing the maximum dimensionless spin magnitude to $0.9$ for compact objects with component masses greater than $0.5\,\Msun$~\citep{LIGOScientific:2021job, LIGOScientific:2022hai, Brown:2012kd}, settings we continue to use in \ac{O4a}.
Although \ac{PBH} and \ac{DBH} formation scenarios can populate the mass spectrum below $0.2\,\Msun$, the target parameter space is restricted to alleviate the computational cost of matched-filtering searches.
Moreover, the horizon distance of \ac{LIGO} and Virgo detectors decreases with decreasing mass~\citep{LIGOScientific:2018glc, Allen:2005fk, Fairhurst:2007qj}.
More recently, searches for \ac{PBH} binaries have been conducted in \ac{O3} and \ac{O4a} data utilizing their quasi-monochromatic evolution in the LIGO-Virgo-KAGRA bandwidth to set limits on the abundance of $[10^{-7},10^{-2}]\,\Msun$ \acp{PBH}~\citep{2024PhRvL.133k1401M, 2025arXiv251119911T}.
Strain upper limits from continuous \ac{GW} searches for non-axisymmetric rotating neutron stars in \ac{O4a} data also constrain \ac{PBH} binaries with chirp masses $\sim 10^{-5}\,\Msun$ since they mimic continuous emission in this frequency band~\citep{LIGOScientific:2026plm}.

The absence of confident detections in previous runs has provided useful constraints on the abundance of \acp{PBH}, and dissipative dark matter models~\citep{LIGOScientific:2022hai}.
Motivated by the \ac{SSM} searches' sensitivity to typical electromagnetically-bright sources, for the first time we operated two independent low-latency \ac{SSM} searches during the \ac{O4} to enable prompt follow-up in the event of a significant candidate.
Details of these efforts are provided in~\cite{Allene:2025saz} and~\cite{Hanna:2024tom}.
The real-time analyses were followed by archival searches, which evaluate candidate significance using the noise background collected across the entire observing run.
In this paper, we focus on results from the archival \ac{SSM} searches performed by three pipelines: \GSTLAL, \MBTA and \PYCBC.
We report their detection sensitivities for \ac{SSM} \acp{BBH} and \acp{BNS} by repeating the searches on strain data from O4a with simulated signals added.
Further, we derive limits on the merger rate of \ac{SSM} \acp{BBH}, which are used to update the constraints on the fraction of dark matter in \acp{PBH} and the fraction of dark matter in \acp{DBH}.
For the \ac{PBH} scenario, we present constraints across the component-mass plane in addition to previously used point-mass distributions~\citep{LIGOScientific:2018glc, LIGOScientific:2019kan, LIGOScientific:2021job, LIGOScientific:2022hai}.

In addition to our constraints on specific dark matter models, we report the sensitivity of the \ac{SSM} search to binaries comprising one or two low-mass \acp{NS}.
\ac{SSM} searches cover the range of \ac{BNS} chirp masses, while extending into the high-spin parameter space typically excluded in template-bank based searches for \ac{BNS}~\citep{LIGOScientific:2025slb}.
Matter effects involving the tidal interaction of the binary components influence binary phasing starting at an effective fifth post-Newtonian order.
For \acp{BNS} with super-solar masses, they affect waveforms at $\geq$ 1\,kHz where \ac{LIGO} and Virgo have reduced sensitivity.
However, extremely large tides at low \ac{NS} masses ($\leq 0.5\,\Msun$) cause dephasing below 1\,kHz which can result in significant mismatches between the \ac{BBH} template waveforms and the \ac{BNS} waveform describing the \ac{GW} signal~\citep{Bandopadhyay:2022tbi}.
Therefore, we extend our sensitivity estimation to include simulated signals from \ac{SSM}~\ac{NS} mergers~\citep{Metzger:2024ujc,Chen:2025uwd} to determine the detection efficiency for this population.

%% file: search.tex
\section{Search}\label{sec:search}


We report results from three independent search pipelines, described in Sections~\ref{sec:gstlal}, \ref{sec:mbta} and~\ref{sec:pycbc}, which analyzed data collected during \ac{O4a}, from \RUNSTART to \RUNEND, and the preceding engineering run, specifically the period of coincident operation of the two \ac{LIGO} detectors from \ERSTART to \EREND.
Details on the detector sensitivity during \ac{O4a} can be found in~\cite{LIGOScientific:2025hdt}, while detector characterization and data quality in \ac{O4a} are described in~\cite{Soni2025} and~\cite{LIGOScientific:2025yae}, respectively.
All three pipelines employ matched-filtering techniques~\citep{LIGOScientific:2025yae}, where the data is correlated with banks of \ac{CBC} waveform models (hereafter referred to as templates) that span a common parameter space across all pipelines, yielding the \ac{SNR} for each candidate event. 
The template banks are designed to search for \acp{CBC} with redshifted primary masses $(1+z)m_1$ between \MOneMIN and \MOneMAX, and redshifted secondary masses $(1+z)m_2$ between \MTwoMIN and \MTwoMAX.
The mass ratio \masstwo/\massone (\masstwo $\leq$ \massone) is restricted within \MassRatioMin--\MassRatioMax to maintain a computationally manageable number of templates in the bank.
We include spin--spin and spin--orbit interaction effects, assuming that the spin vectors of the two bodies are aligned or anti-aligned with the orbital angular momentum vector. 
The dimensionless spin magnitudes satisfy $\spincomponent \leq \MaxSpinHighMass$ for $(1+z)\masscomponent \geq \SSMLowHighMassGap\,\Msun$ and $\spincomponent \leq \MaxSpinLowMass$ for $(1+z)\masscomponent \leq \SSMLowHighMassGap\,\Msun$.
This search covers the same parameter space as the \ac{LVK} \ac{SSM} search performed on \ac{O3} data, making a trade-off between possible astrophysical targets and the computational cost of matched-filtering searches in this low-mass region.
All pipelines design template banks with a minimal match of $\sim \MinMatch$, ensuring that no more than $10\%$ of astrophysical signals are lost due to the discretization of the parameter space~\citep{Owen:1995tm}.
Although the sensitivity bands of \ac{LIGO} and Virgo extend down to approximately 15\,Hz~\citep{LIGOScientific:2025hdt}, our templates are truncated to start at higher frequencies, chosen independently by each pipeline, to reduce the computational cost, since \ac{GW} signals from low-mass \acp{CBC} can last for several minutes within the detector band.
Details about the \ac{SSM} template bank generation, search optimization including low-frequency cutoffs, and waveform models specific to each pipeline are provided in the following subsections.

\subsection{\GSTLAL}\label{sec:gstlal}

\GSTLAL~\citep{Cannon:2011rj, Messick:2016aqy, Sachdev:2019vvd, Hanna:2019ezx, Cannon:2020qnf, Sakon:2022ibh, Tsukada:2023edh} is a GStreamer-based pipeline that searches for \ac{GW} signals from \acp{CBC} using matched-filtering techniques.
A concise overview of the methods developed and adopted for \ac{O4} analyses, including the latest archival search workflow employed in this work, is provided in~\cite{LIGOScientific:2025yae, Joshi:2025nty, Joshi:2025zdu}.
Unlike in \ac{O3}~\citep{LIGOScientific:2021job, LIGOScientific:2022hai}, \GSTLAL uses a bin-dependent gating procedure to mitigate glitches instead of the iDQ algorithm~\citep{Huxford:2024vdt}.

In contrast to previous \ac{SSM} searches, \GSTLAL generated the \ac{SSM} template bank using a geometric placement method \MANIFOLD~\citep{Hanna:2022zpk}.
This \ac{SSM} template bank spans the parameter space described above but is constructed with a slightly lower minimum match of \GSTLALSSMMinMatch~\citep{Hanna:2024tom} to limit its size.
The resulting bank comprises over $3.0$ million templates~\citep{Hanna:2024tom}.
Templates are modeled using \IMRPHENOMD~\citep{Khan:2016, Husa:2016} with a low-frequency cutoff of at least \GstlalFLOW and a high-frequency cutoff of \GstlalFHIGH.
Template waveforms are restricted to a maximum duration of approximately \GstlalMAXDUR prior to merger, which may truncate the waveform relative to one starting at 45\,Hz.
A detailed description of the geometric placement method, as well as the design and performance of the bank, is given in~\cite{Hanna:2024tom}. 

The current \ac{SSM} search incorporates a number of enhancements implemented in \GSTLAL analyses during \ac{O4} compared to the \ac{O3} analysis~\citep{KAGRA:2021vkt}.
These developments were introduced for the GWTC-4.0 searches~\citep{LIGOScientific:2025yae} and are likewise employed in the \ac{SSM} search, which include an improved likelihood function~\citep{Tsukada:2023edh}, a new \ac{SVD}-based grouping method for parallelization~\citep{Sakon:2022ibh}, and a template weighting derived from the \MANIFOLD\ mass model~\citep{Ray:2023nhx, LIGOScientific:2025yae}. 
As in the \ac{O3} \ac{SSM} search, we continue to assume a uniform population model in the ranking statistic due to the absence of observational constraints on the underlying source population.
The data are match-filtered using the same frequency and waveform duration cutoffs applied in the template bank generation, with waveform models chosen based on redshifted chirp mass $(1+z)\Mc = (1+z)(\massone \masstwo)^{3/5}/(\massone + \masstwo)^{1/5}$: the \TAYLORFTWO approximant~\citep{Buonanno:2009ab, Pan:2008pd, Boyle:2009bb} is used for $(1+z)\Mc \leq 1.73\,\Msun$, and the \SEOBNRVFOURROM approximant~\citep{Buonanno:2009ab, Bohé:2017aa, Cotesta:2020ed} for $(1+z)\Mc > 1.73\,\Msun$.
We adopt different waveform models from that used for template placement in order to reduce computational costs.
After matched-filtering, triggers---potential \ac{GW} signals in data---are ranked using a likelihood function~\citep{Tsukada:2023edh, Cannon:2015gha}.
The likelihood term for multi-detector signal consistency~\citep{Hanna:2019ezx} is evaluated over a frequency range of \GstlalFLOW to \GstlalDtDphiFHIGH for this search.
Although this high-frequency cutoff is lower than \GstlalFHIGH used in template whitening, we find negligible difference in signal weighting between cutoffs of \GstlalDtDphiFHIGH and \GstlalFHIGH for this likelihood term.
Ultimately, all triggers are assigned a \ac{FAR} which quantifies the significance of a candidate signal by comparing it to the distribution of likelihoods for non-astrophysical events~\citep{Messick:2016aqy}.

\subsection{\MBTA}\label{sec:mbta}

The Multi-Band Template Analysis (\MBTA; \citealp{LIGOScientific:2012fcp, Adams:2015ulm, Aubin:2020goo, Allene:2025saz}) is a matched-filtering-based pipeline that searches for \acp{GW} from \acp{CBC} in data from the LIGO--Virgo--KAGRA detectors.
\MBTA stands out by its approach of splitting the matched-filtering process across multiple frequency bands to optimize computational efficiency.
This approach significantly reduces the computational cost of processing a large template bank compared with an analysis performed over a single frequency band~\citep{Allene:2025saz}.

The template bank used by \MBTA is the same as the one employed in the low-latency analysis described in~\cite{Allene:2025saz}.
It comprises more than $2.5$ million templates, covering the parameter space described above, and was generated using a geometric placement algorithm~\citep{Brown:2012nn}.
The waveforms used to build the template bank are computed using the second-order post-Newtonian~\TAYLORFTWO~\citep{Buonanno:2009ab, Pan:2008pd, Boyle:2009bb}, while the \MBTASEARCHAPPROX approximant~\citep{Isoyama:2020lls} is used for the analysis.
A low-frequency cutoff of \MBTAFLOW and high-frequency cutoff of \MBTAFHIGH are applied to limit the duration of the templates, ensuring that the filtering remains computationally manageable while preserving the search sensitivity~\citep{Allene:2025saz}.

As an improvement over \ac{O3}, \MBTA now includes a single-detector trigger search in its archival analysis~\citep{Allene:2025saz}, enabling the use of periods when only one of the two \ac{LIGO} detectors is operational.
The post-processing procedure has also been refined with the implementation of a new internal data quality assessment tool called the SNR-Excess technique which is detailed in~\cite{Allene:2025saz}.

\subsection{\PYCBC}\label{sec:pycbc}

The \PYCBC pipeline~\citep{Allen:2005fk, Canton:2014ena, Usman:2015kfa, Nitz:2017svb, Davies:2020tsx, pycbcsoft} performs a matched-filtering analysis to detect \ac{GW} signals from \acp{CBC}.
The \PYCBC \ac{SSM} search employs the latest version of the pipeline, described in detail in~\cite{LIGOScientific:2025yae}.
However, specific choices were made here to tailor the analysis to the search for \ac{SSM} compact objects and improve its sensitivity.
These changes are summarized below.

The ranking statistic is the same as in the \ac{O3} \ac{SSM} search~\citep{LIGOScientific:2021job}, with the only difference being that corrections for short term variations of the \ac{PSD}~\citep{Mozzon:2020gwa} are not included.
The \PYCBC \ac{SSM} template bank contains approximately $2.0$ million templates, and was generated using a geometric placement algorithm~\citep{Brown:2012nn}.
The \TAYLORFTWO post-Newtonian approximant~\citep{Buonanno:2009ab, Pan:2008pd, Boyle:2009bb}, a low-frequency cutoff of \PyCBCBankFLOW, and a high-frequency cutoff of \PyCBCBankFHIGH were adopted to compute the matches between templates.

The pipeline starts filtering the data at \PyCBCFilterFLOW, with a maximum allowed template duration of \PyCBCMaxTempltDur imposed to keep computational resources usage under control. 
The limit on the duration affects only about 6$\%$ of the templates, those with the lowest masses, increasing their starting frequency to approximately 33\,Hz at most.
Adopting a starting frequency lower than that used for the template bank generation implies that the design minimum match of \MinMatch is no longer strictly guaranteed for this filtering configuration. 
However, this choice was found to improve the recovered \ac{SNR} of signals compared to the more straightforward approach of starting the filtering from \PyCBCBankFLOW.
By contrast, using \PyCBCFilterFLOW as the low-frequency cutoff for bank generation would have required a substantially larger number of templates, making the search computationally impractical.
The \SEOBNRVFOURROM approximant~\citep{Buonanno:2009ab, Bohé:2017aa, Cotesta:2020ed} is used instead of \TAYLORFTWO for waveform generation when the redshifted total mass $(1+z)(\massone + \masstwo) > \PyCBCMtotThr$.

Analogously to MBTA, starting in \ac{O4}, \PYCBC includes single-detector triggers in this archival analysis following~\cite{Davies:2022thw}. This increases overall sensitivity, and allows for detection of events occurring when only one detector is observing.

\subsection{Results}\label{sec:pipeline_result}
\begin{figure*}
    \centering
    \includegraphics[width=1.0\textwidth]{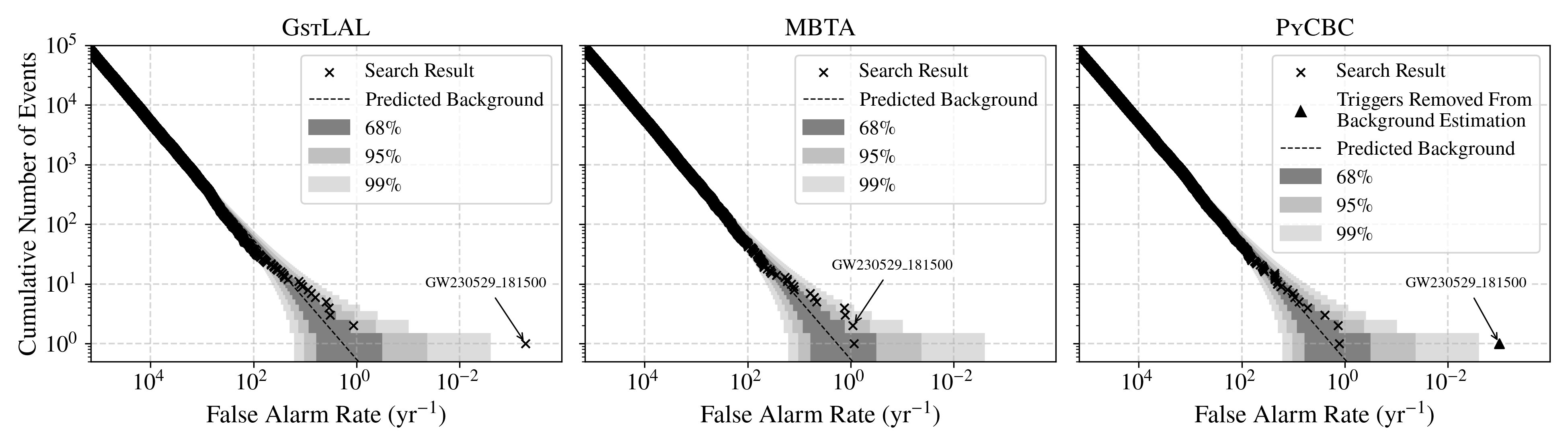}
    \caption{Cumulative number of \ac{SSM} candidate events as a function of the \ac{FAR} (in yr$^{-1}$) for the \GSTLAL (left panel), \MBTA (middle panel), and \PYCBC (right panel) analyses.
             The black dashed curve shows the predicted background with the shaded regions indicating the 68\%, 95\%, and 99\% confidence intervals for the expected background count at each \ac{FAR} threshold, derived from a Poisson distribution given the total analyzed time of \ac{O4a}.
             The observed cumulative count from each pipeline is shown as black crosses at each threshold.
             The high-significance trigger seen in the \GSTLAL and \PYCBC distributions corresponds to GW230529\_181500 from GWTC-4.0~\citep{LIGOScientific:2024elc}, whose inferred secondary mass exceeds one solar mass. Therefore, it is not considered an \ac{SSM} candidate.
	     In the \PYCBC plot, GW230529\_181500 is marked with a black triangle indicating that it is not considered in the cumulative count because the analysis removes events with \acp{FAR} $\leq$ 1 per $100\,\rm{yr}$ to reduce contamination from loud signals during background estimation for less significant triggers~\citep{LIGOScientific:2016dsl,Nitz:2018imz}.
No other deviation from the expected background is observed, indicating we find no significant \ac{GW} transient from \ac{SSM} candidates.}
    \label{fig:search_results_FAR_plots}
\end{figure*}

We report no significant \ac{GW} transients from \ac{SSM} candidates in \GSTLAL, \MBTA and \PYCBC search analyses conducted on \ac{O4a} data.
All observed triggers are statistically consistent with a Poisson distribution of noise background events, except for a single outlier corresponding to the published event GW230529\_181500~\citep{LIGOScientific:2024elc}. 
This outlier, identified as the most significant trigger by both \GSTLAL and \PYCBC, is a single-detector event observed at the \ac{LLO}.
GW230529\_181500 is the second-most significant trigger in the \MBTA \ac{SSM} search because it is down-ranked by \MBTA's more conservative significance assignment for single-detector events.
The secondary mass of this trigger is inferred to exceed one solar mass~\citep{LIGOScientific:2024elc}, indicating that it does not contain an \ac{SSM} candidate.
Figure~\ref{fig:search_results_FAR_plots} presents the results from \GSTLAL~(left panel), \MBTA~(middle panel), and \PYCBC~(right panel), showing the number of events with assigned significance above a given \ac{FAR} threshold as a function of the threshold.

Table~\ref{tab:triggers} lists triggers from \GSTLAL, \MBTA, and \PYCBC with a \ac{FAR} below \SignificantOPA, including those associated with GW230529\_181500.
The reported \acp{FAR} do not include any trials factor.
The most significant \MBTA trigger is consistent with the expected background and shows no coincidence with any significant triggers from other pipelines.
The second most significant \GSTLAL trigger, listed fifth in Table~\ref{tab:triggers}, was coincidentally recovered with the same $(1+z)\massone$ and $(1+z)\masstwo$ as GW230529\_181500 at \GSTLALSecondTriggerTime, but appears to be consistent with the background distribution.
\MBTA and \PYCBC recover this trigger with \acp{FAR} well above the \SignificantOPA, and GWTC-4.0 reports it as GW230528\_110748 with even lower significance~\citep{LIGOScientific:2025slb}.

Considering that the only significant trigger corresponds to GW230529\_181500~\citep{LIGOScientific:2024elc} and that the second most significant \GSTLAL trigger is consistent with noise, we conclude that no significant \ac{GW} transient from \ac{SSM} candidates are observed.

\begin{table*}[ht]
  \tablewidth{\linewidth}
  \setlength{\tabcolsep}{3pt} 
  \renewcommand{\arraystretch}{1.2} 
  \caption{\label{tab:triggers}Triggers recovered with a \ac{FAR} below \SignificantOPA by \GSTLAL, \MBTA and \PYCBC.}
  \begin{tabular}{ c  c  c  c  c  c  c  c  c }
    \hline \hline
    \ac{FAR} & Pipeline & UTC & $(1+z)\massone$ & $(1+z)\masstwo$ & \spinone & \spintwo & Inst. & Network \ac{SNR} \\
    ($\mathrm{yr^{-1}}$) &&& (\Msun) & (\Msun) &&&& \\
    \hline
    \expandafter\GSTLALTableRow\GSTLALTrigOneData
    \expandafter\PYCBCTableRow\PYCBCTrigOneData
    \expandafter\MBTATableRow\MBTATrigOneData
    \expandafter\MBTATableRow\MBTATrigTwoData
    \expandafter\GSTLALTableRow\GSTLALTrigTwoData
    \expandafter\PYCBCTableRow\PYCBCTrigTwoData
    \expandafter\MBTATableRow\MBTATrigThreeData
    \expandafter\MBTATableRow\MBTATrigFourData
    \expandafter\PYCBCTableRow\PYCBCTrigThreeData
    \hline
  \end{tabular}
\tablecomments{We report the redshifted masses $(1+z)\masscomponent$ and the dimensionless spin magnitudes \spincomponent of the primary (1) and secondary (2) components corresponding to the template that yielded the highest ranking statistic for each trigger.
  The Network \ac{SNR} is defined as the square root of the quadrature sum of the \ac{SNR} measured in all detectors operating at the time of the trigger.
  The detectors which were observing at the time of each event are listed under the `Inst.' column, denoted by H for LIGO Hanford and L for LIGO Livingston.
  Triggers corresponding to GW230529\_181500~\citep{LIGOScientific:2024elc} in results from \GSTLAL, \PYCBC, and \MBTA appear as the first, second, and fourth entries respectively.
}
\end{table*}

%% file: search_sensitivity.tex
\section{Search Sensitivity}\label{sec:VT}

We quantify the sensitivity of our searches with the \ac{O4a} detector network to \ac{GW} signals from \ac{SSM} \acp{CBC} to set an upper limit on their merger rates.
Due to the changing sensitivity across the observing run and the complex nature of the detection process, we determine this sensitivity empirically via the procedure laid out in~\cite{LIGOScientific:2025yae, LIGOScientific:2025pvj, Essick:2025zed}. 
Following~\cite{Essick:2025zed}, we simulate and inject signals into detector strain data, and all pipelines perform the search described in Section~\ref{sec:search} on this augmented data.
Any statement of our sensitivity to \ac{SSM} sources depends on the assumed astrophysical population that describes their distribution of masses and spins. 
Therefore, we state the sensitivity to \ac{SSM} sources \textit{as a function of an assumed population}.
This allows us to use the results from a single injection-recovery analysis to determine the sensitivity to multiple population models.

The sensitive space-time hypervolume $\langle VT \rangle$, and its corresponding uncertainty $\sigma_{\langle VT \rangle}$ for an assumed population denoted by $\mathbf{\Lambda}$ is given by
\begin{align}\label{eq:VT_from_xi}
    \langle VT \rangle(\mathbf{\Lambda}) = \mathcal{\xi}(\mathbf{\Lambda}) VT_{\rm total}, \\
    \sigma_{\langle VT \rangle} = VT_{\rm total} {\sigma_{\mathcal{\xi}}}, \\
    VT_{\rm total} = T_{\rm obs} \int^{z_{\rm max}}_0 \mathrm{d}z \frac{1}{1+z} \frac{\mathrm{d}V_c}{\mathrm{d}z},
\end{align}
where $\xi(\mathbf{\Lambda})$ is the fraction of detectable mergers which depends on a chosen detection threshold, $VT_{\rm total}$ is the hypervolume spanned by the injected population assuming a constant number of events per unit time per unit comoving volume within a sphere of redshift $z_{\rm max}$, $\sigma_{\mathcal{\xi}}$ is the square root of the variance in the estimate of $\mathcal{\xi}$, and $T_{\rm obs}$ denotes the time spanned by the augmented data containing simulated signals, which in this case is the duration of \ac{O4a}.

We consider two fiducial populations from which simulated sources are drawn to assess the sensitivity of the search: a \ac{BBH} population with at least one \ac{SSM} component, and a \ac{BNS} population, which includes tidal interaction effects, extending into the sub-solar regime.
The \ac{BBH} population is largely similar to previously considered fiducial populations~\citep{LIGOScientific:2022hai}, but an \ac{SSM} \ac{BNS} population has not been considered in \ac{LVK} analyses previously.
If \ac{SSM} \ac{BNS} systems exist, their \ac{GW} signals will exhibit effects due to tidal interactions.
These effects are significantly more pronounced for low-mass \acp{NS}, so the sensitivity of the \ac{SSM} search will depend on the composition of the assumed population.
We model the \ac{BNS} signals using the \IMRPHENOMXPNRTIDALVTHREE waveform model~\citep{Pratten:2020ceb, Abac:2023ujg} with tidal effects given by the \GPPVADDtwo equation of state~\citep{Servignat:2023lai} as implemented in the \textsc{LALSuite} library for \ac{GW} data analysis~\citep{lalsuite}.
Tidal effects for systems including low-mass \acp{NS} are extrapolated from those found in numerical simulations of systems with components above $1\,\Msun$.
However, this equation of state produces large-radius stars with tides near the upper limit allowed by previous \ac{GW} observations~\citep{LIGOScientific:2017vwq, Radice:2017lry}, so should give a reasonably conservative estimate for our sensitivity to \ac{SSM} \acp{BNS} using template waveforms that do not include tides.

Table~\ref{tab:injection_population} summarizes the properties of the \ac{BBH} and \ac{BNS} fiducial populations ($\mathbf{\Lambda}_0$).
The lowest component mass for the \ac{BBH} population is chosen in accordance with the search parameter space which extends to \MOneMIN.
For \acp{BNS}, we restrict component masses to be greater than \MinPrimaryMassNS.
At \MinPrimaryMassNS, tidal effects extrapolated for the \GPPVADDtwo equation of state are on the order of $10^5$ and would be even higher at lower \ac{NS} masses, thereby exacerbating systematic uncertainties from waveform validity.
The secondary component mass is restricted such that every source has at least one sub-solar mass component.
For both populations, we choose an equally weighted mixture model of a half normal distribution and a Beta distribution such that the dimensionless spin magnitude peaks at \spincomponent $=$ \MinSpinMag~with small support extending to high spins.
This choice is motivated by expected low spins for primordial black holes~\citep{Garcia-Bellido:2017fdg, Chiba:2017rvs, DeLuca:2019buf,DeLuca:2020bjf}, and observed low spins for Galactic BNS~\citep{Stovall:2018ouw} without excluding support for highly-spinning systems entirely.
The component spins are isotropically distributed, i.e. the cosine of the angles between the orbital angular momentum vector and the component spin vectors are distributed uniformly.
The sources are distributed uniformly in comoving volume up to a maximum redshift $z_{\rm max} =$ \MaxRedshiftInj~which results in a total hypervolume $VT_{\rm total} =$ \TotalInjVT.

Details on the calculation of the fraction of detectable mergers, $\mathcal{\xi}(\mathbf{\Lambda})$ using searches performed on the fiducial injection population ($\mathbf{\Lambda}_0$) are provided in Appendix~\ref{appendix:vt}.
The choice of \ac{FAR} thresholds for estimating $\mathcal{\xi}(\mathbf{\Lambda})$ are described in relevant sections.
We use the software package \GWPOPULATION~\citep{Talbot2025} to compute the fraction of detectable mergers $\xi(\mathbf{\Lambda})$ through importance sampling (Equation~\ref{eq:MC_efficiency}), and finally compute the sensitive hypervolume $\langle VT \rangle$ (Equation~\ref{eq:VT_from_xi}).
The following subsections detail the $\langle VT \rangle$ for selected \ac{BBH} and \ac{BNS} models.
Additionally, we provide the complete injection datasets and search results in a supplemental data release~\citep{ZenodoO4aSSM}, allowing for the reweighting of these fiducial results to alternative populations.

\subsection{Sub-solar mass \ac{BBH}}\label{sec:bbh_vt}
We report the $\langle VT \rangle$ for \ac{BBH} systems with at least one component less massive than 1\,\Msun as a function of the chirp mass \Mc of the binary.
The fiducial population ($\mathbf{\Lambda}_0$; Table~\ref{tab:injection_population}) which is log-uniform in component masses, is re-weighted in its spin distribution to be isotropic and with uniform component spin magnitudes $\spincomponent \leq \MaxSpinMagVTpop$.
Additionally, the mass ratio is restricted to $\masstwo / \massone \geq 0.05$ to capture a broader astrophysical population than covered by the template banks, while still excluding extreme mass ratios.
The reweighted distributions match the \ac{BBH} distribution considered in~\cite{LIGOScientific:2022hai} and ensure that $\langle VT \rangle$ from \ac{O4a} is directly comparable with $\langle VT \rangle$ from \ac{O3}.

We use the \ac{FAR} of the most significant noise candidate from \ac{O4a} data as the detection threshold to calculate $\langle VT \rangle$, allowing for the calculation of upper limits on the merger rate through the loudest event statistic formalism~\citep{Biswas:2007ni}.
Specifically, we assume \ac{FAR} thresholds of \GSTLALLoudestTriggerFAR, \MBTALoudestTriggerFAR~and \PYCBCLoudestTriggerFAR~for \GSTLAL, \MBTA and \PYCBC respectively.

\begin{figure}[htb]
    \centering
    \includegraphics[width=\linewidth]{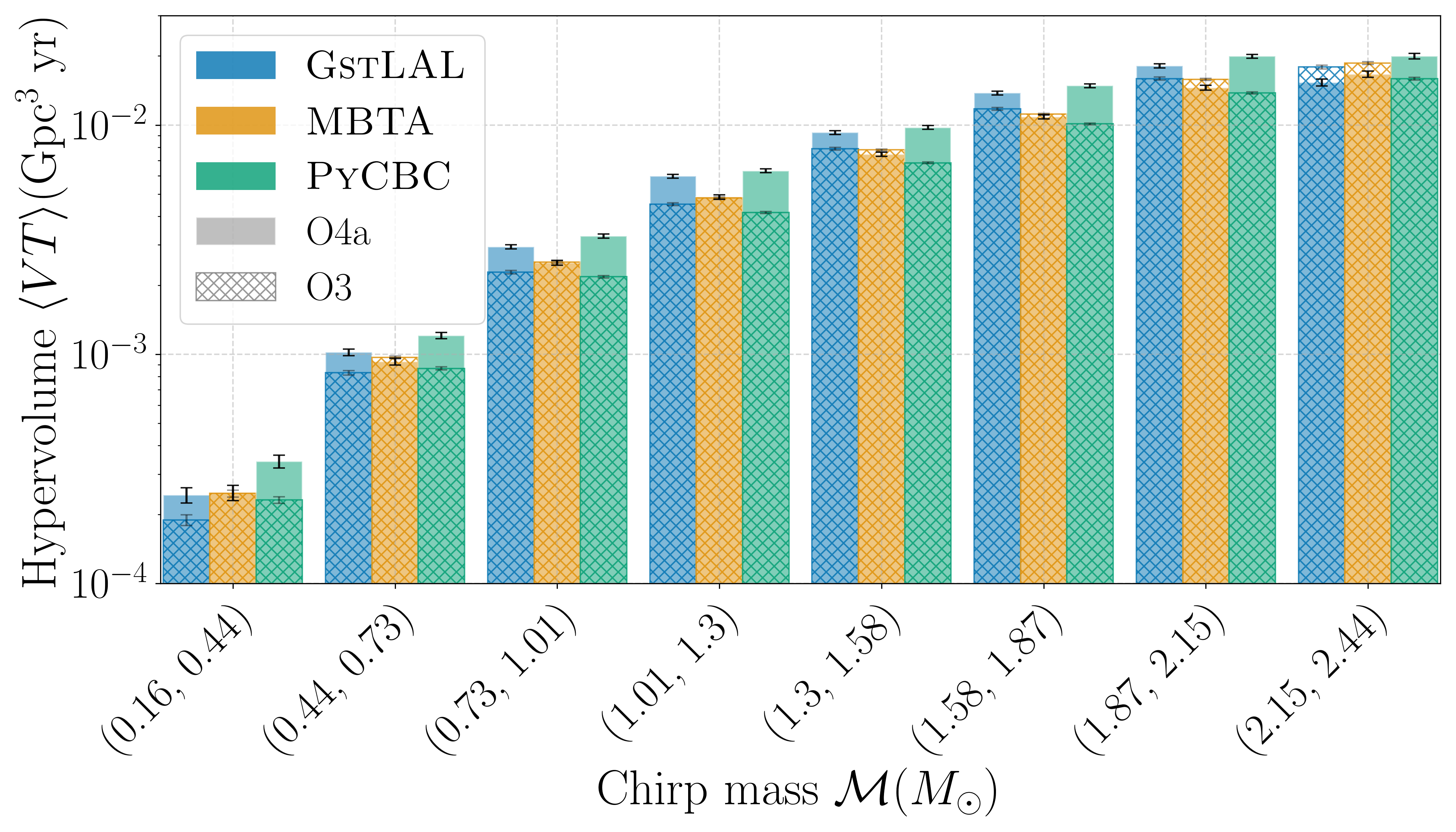}
    \caption{Sensitive hypervolume, $\langle VT \rangle~(\mathrm{Gpc^3\,yr})$ as a function of the chirp mass, \Mc for searches \GSTLAL (\textit{blue}), \MBTA (\textit{yellow}) and \PYCBC (\textit{green}). The masses of the detected sources within each \Mc bin are reweighted to a log-uniform distribution with $\MinSecondaryMassBH \leq \masstwo \leq \MaxSecondaryMassBH$ and $\masstwo \leq \massone \leq \MaxPrimaryMassBH$ with $q \geq 0.05$, and their spin distribution is reweighted to an isotropic uniform distribution with $\spincomponent \leq \MaxSpinMagVTpop$ to estimate $\langle VT \rangle_{\rm O4a}$ (\textit{solid bars}). $\langle VT \rangle_{\rm O3}$ (\textit{hatched bars}) for each search are shown for reference~\citep{LIGOScientific:2022hai}. Error-bars reflect the $1\sigma$ uncertainty in $\langle VT \rangle$.}
    \label{fig:VT_mc}
\end{figure}

The total $\langle VT \rangle$ of our search is dominated by the more massive systems in the population model.
In order to provide a more refined estimate of our sensitivity across mass space, and in order to compare to previously reported results from \ac{O3}~\citep{LIGOScientific:2022hai}, we estimate our sensitivity in individual mass bins.
Such bin-wise estimates also partially mitigate, but do not remove entirely, the systematic uncertainty in our $\langle VT \rangle$ estimate due to having to choose a population model.
Figure~\ref{fig:VT_mc} shows the $\langle VT \rangle$ computed over \ac{O4a} as a function of \Mc for the three searches separately, with $\langle VT \rangle_{\rm O3}$ plotted for comparison.
We divide the range of chirp masses into 8 linearly spaced bins between $0.16\,\Msun$ to $2.44\,\Msun$.
The simulated signals which pass the detection threshold are reweighted to the desired population ($\mathbf{\Lambda}$) described above using Equation~(\ref{eq:MC_efficiency}) to compute the $\langle VT \rangle$ within each \Mc bin.
The $\langle VT \rangle_{\rm O4a}$ for the three searches remains similar to the $\langle VT \rangle_{\rm O3}$, as the expected gain from improved sensitivity of the detectors and searches in \ac{O4} is offset by the shorter observing period of \FourthAObsAnalysisTime~in \ac{O4a} compared to \ThirdObsAnalysisTime~in \ac{O3}.
The mean ratio of $\langle VT\rangle_{\rm O4a} $ to $\langle VT\rangle_{\rm O3} $ across all \Mc bins is \GstLALVTMcratio, \MBTAVTMcratio, and \PyCBCVTMcratio~for \GSTLAL, \MBTA, and \PYCBC respectively.
The appreciable increase for \PYCBC is partly due to to the reduced low-frequency cutoff of 30\,Hz in the \ac{O4a} analysis.
The reduced sensitivity of the \GSTLAL search in the most massive bin can be attributed to the slightly worse performance of the template bank in the high mass region(see Figure 6 in~\cite{Hanna:2024tom}).
While bank simulations confirm that the bank density remains sufficient to ensure optimal coverage~\citep{Hanna:2024tom}, the higher mismatches in this region are reflected in $\langle VT \rangle$ when evaluated as a function of \Mc.
For \MBTA, the reduced $\langle VT\rangle_{\rm O4a}$ in this bin is likely due to the greater mismatch between the injected \IMRPHENOMXPHM and template \MBTASEARCHAPPROX waveforms at high masses, unlike in \ac{O3} where the injected and template waveforms were identical.

\begin{figure}[h]
    \centering
        \includegraphics[width=\linewidth]{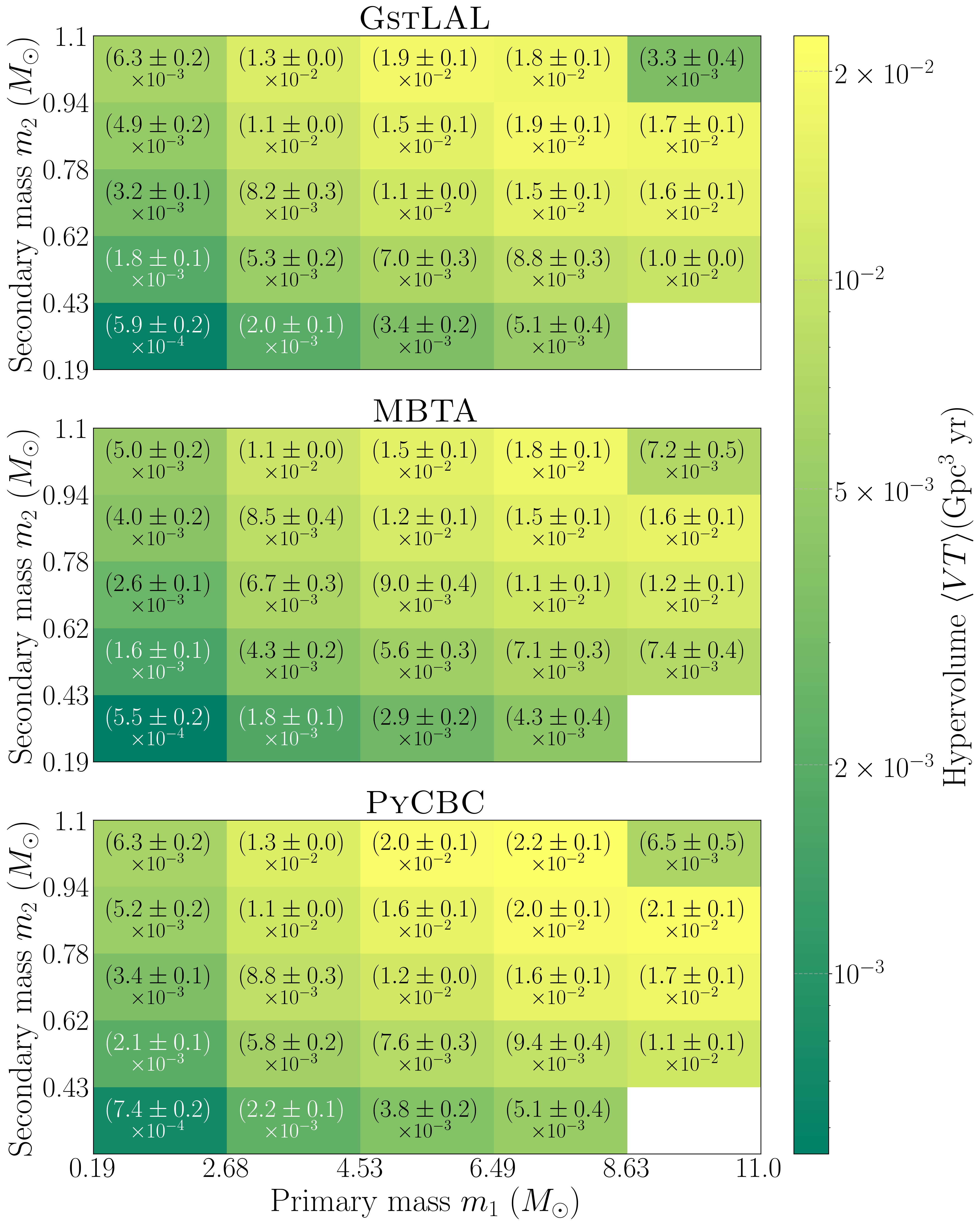}
    \caption{The sensitive hypervolume, $\langle VT \rangle$ as a function of the component masses, (\massone, \masstwo), computed for searches \GSTLAL, \MBTA and \PYCBC following the injection campaign in data from \ac{O4a}. The color of each cell corresponds to the $\langle VT \rangle_{\rm O4a}$ value. The sources within each bin are reweighted to a log-uniform distribution with $\MinSecondaryMassBH \leq \masstwo \leq \MaxSecondaryMassBH$ and $\masstwo \leq \massone \leq \MaxPrimaryMassBH$ with $q \geq 0.05$, and the spin distribution is reweighted to an isotropic uniform distribution with $\spincomponent \leq \MaxSpinMagVTpop$ to estimate $\langle VT \rangle_{\rm O4a}$. Errors represent the $1\sigma$ uncertainty in $\langle VT \rangle$.}
    \label{fig:VT_m1m2}
\end{figure}

We also report $\langle VT \rangle_{\rm O4a}$ as a function of component masses, \masscomponent, shown in Figure~\ref{fig:VT_m1m2} for the same population, and the same per-search detection threshold.
We divide the range of component masses, \massone (\masstwo) into 5 linearly spaced bins between $0.19\,\Msun$ to $11.0~(1.1)\,\Msun$.
The detected simulated signals are then reweighted to the desired population $\mathbf{\Lambda}$ using Equation~(\ref{eq:MC_efficiency}) to compute the $\langle VT \rangle$ within each bin.
We observe similar trends in this case for the individual searches: the ratio of $\langle VT \rangle_{\rm O4a}$ to $\langle VT \rangle_{\rm O3}$ averaged across all bins is \GstLALVTCompMassratio, \MBTAVTCompMassratio, and \PyCBCVTCompMassratio~for \GSTLAL, \MBTA, and \PYCBC respectively.
Expectedly, the sensitivity increases for more massive binaries, while we find reduced sensitivity for more asymmetric systems.

\subsection{Sub-solar mass \ac{BNS}}

\begin{figure}[t]
    \centering
        \includegraphics[width=\linewidth]{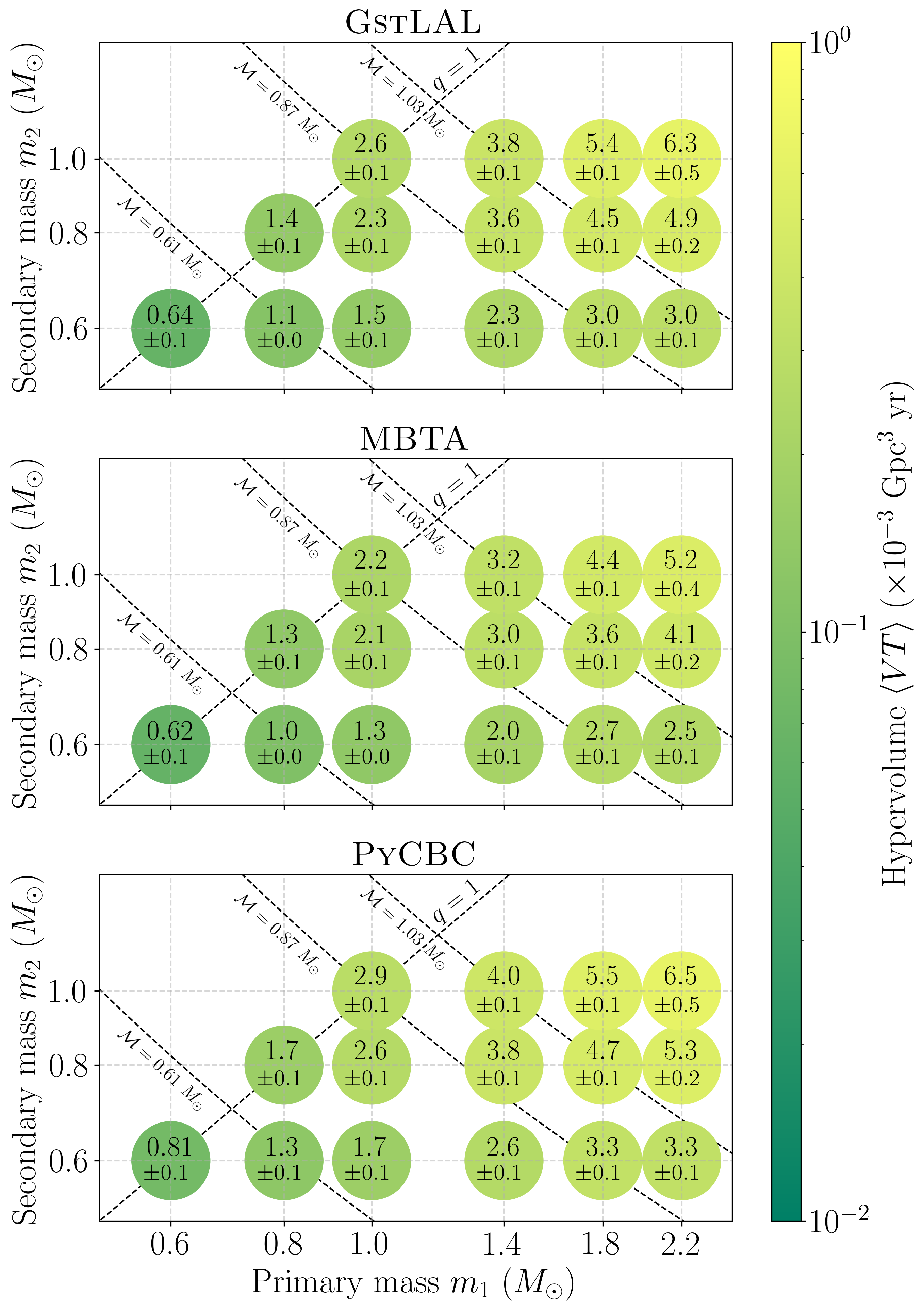}
\caption{The sensitive hypervolume, $\langle VT \rangle~(\mathrm{Gpc^3\,yr})$ for a high-spinning \ac{BNS} population applying a significance threshold \ac{FAR} $\leq$ \BNSVTFARThreshold. We report the $\langle VT \rangle_{\rm O4a}$ for \GSTLAL, \MBTA and \PYCBC computed using Equation~(\ref{eq:VT_from_xi}). The color of each circle corresponds to the $\langle VT \rangle_{\rm O4a}$ value. The plotted points correspond to the central points of log-normal distributions with widths \LogNormSigmaBNSVT~used to estimate $\langle VT \rangle_{\rm O4a}$. Errors represent the $1\sigma$ uncertainty in $\langle VT \rangle$.}
\label{fig:bns_vt}
\end{figure}

Following the search sensitivity as reported in~\cite{LIGOScientific:2025slb}, we compute the sensitive hypervolume for \ac{BNS} systems at representative combinations of component masses for two distinct cases extending into the low-mass \ac{NS} parameter space:
\begin{itemize}
    \item Double \ac{SSM}: \masscomponent $\leq 1\,\Msun$. We pick three combinations with \masscomponent equal to 0.6\,\Msun, 0.8\,\Msun, and 1\,\Msun with $\masstwo \leq \massone$.
    \item Single \ac{SSM}: \massone $> 1\,\Msun$ while \masstwo $\leq 1\,\Msun$. We pick combinations of \massone equal to 1.4\,\Msun, 1.8\,\Msun, and 2.2\,\Msun with \masstwo equal to 0.6\,\Msun, 0.8\,\Msun, and 1\,\Msun.
\end{itemize}
The primary difference in the considered \ac{BNS} populations compared to analyses reported in~\cite{LIGOScientific:2025slb, LIGOScientific:2025pvj} is the extension of the \ac{NS} mass distributions down to \MinPrimaryMassNS.
Unlike the \ac{BBH} case, we compute the sensitive hypervolume at each point by reweighting the injections to a log-normal distribution about the central mass with a width of \LogNormSigmaBNSVT.
Therefore, 0.6\,\Msun is the lowest central mass we consider to avoid regions of the parameter space with no injections.
For each point in the component mass parameter space, we assume an isotropic distribution of spins with spin magnitudes \spincomponent $\leq$ \maxspinBNS.
Figure~\ref{fig:bns_vt} shows the resulting variation in $\langle VT \rangle_{\rm O4a}$ with a detection threshold of \ac{FAR} $\leq$ \BNSVTFARThreshold~for each search following the detection threshold used for computing $\langle VT \rangle$ in~\cite{LIGOScientific:2025slb}.

We present the sensitivity of \ac{SSM} searches toward \ac{BNS} mergers using point-mass estimates to demonstrate how they complement standard GWTC searches~\citep{LIGOScientific:2025slb}.
On reweighting the \ac{BBH} population identically as the \ac{BNS} population, we find that $\langle VT \rangle_{BNS}$/$\langle VT \rangle_{BBH}$ at $(0.6, 0.6)\,\Msun$ is 0.68 to 0.77 across the three searches. 
We find that the $\langle VT \rangle_{BNS}$ is lower than $\langle VT \rangle_{BBH}$ for systems with at least one \ac{SSM} \ac{NS} driven by the emergence of strong tidal effects at these low masses.
Since we assume a stiff equation of state for the simulated waveforms which result in higher mismatches between \ac{NS} waveform and \ac{BH} template at sub-solar masses~\citep{Bandopadhyay:2022tbi}, the difference in sensitivity is expected to be smaller if the true equation of state is softer.
We also find that the search sensitivity is largely agnostic to different mass models for \ac{NS} like the \textsc{PEAK} and \textsc{POWER} models constrained using data from GWTC-4.0~\citep{LIGOScientific:2025pvj} with $\langle VT \rangle_{\rm O4a}$ comparable to the estimates shown in Figure~\ref{fig:bns_vt}.

\bigskip 
\noindent 
The $1\sigma$ statistical uncertainty on $\langle VT \rangle$ arising from the finite number of samples used to estimate $\mathcal{\xi}(\mathbf{\Lambda})$ (Equation 2) ranges between $2\%$ to $8\%$ for \acp{BBH} and $2\%$ to $10\%$ for \acp{BNS}.
Sources of systematic uncertainty include but are not limited to calibration errors and waveform systematics.
The median calibration error in \ac{O4a} is estimated to be at the percent level~\citep{LIGOScientific:2025snk} resulting in systematic uncertainties $ < 10\%$ in $\langle VT \rangle$.
Waveform systematics from extrapolation of tidal effects for sub-solar mass \ac{BNS} are not fully understood in this sub-solar mass regime. 
Further, numerical relativity simulations used to calibrate waveform models including tidal effects do not cover the low masses and large radii relevant to this regime.
For considered \ac{BBH}, waveform systematics are sub-dominant to statistical errors.
Given the bandwidth of present searches largely covers the inspiral, effects from modeling uncertainties at merger and ringdown are expected to be minimal.

%% file: rates.tex
\section{Rates}\label{sec:rates}

Assuming a Poisson rate for triggers produced by the searches, we use the loudest event statistic formalism~\citep{Brady:2004gt,Biswas:2007ni} to derive the frequentist upper limit on the merger rate for the population of \ac{SSM} \acp{BBH} considered in this work.
Since the most significant events in the three participating search pipelines are consistent with background with the exception of GW230529\_181500 which is conclusively known to have super-solar masses~\citep{LIGOScientific:2024elc}, we derive the upper limit on the merger rate at the 90\% confidence level, $\mathcal{R}_{90,i}$ using $\langle VT\rangle_{\rm O4a}$ as follows:
\begin{equation}\label{eq:rate_limit}
    \mathcal{R}_{90,i} = \frac{2.3}{\langle VT \rangle_i},
\end{equation}
where the $\mathcal{R}_{90, i}$ is computed in the corresponding $i$--th mass bin~\citep{Brady:2004gt,Biswas:2007ni}.

\begin{figure}[b]
    \centering
    \includegraphics[width=\linewidth]{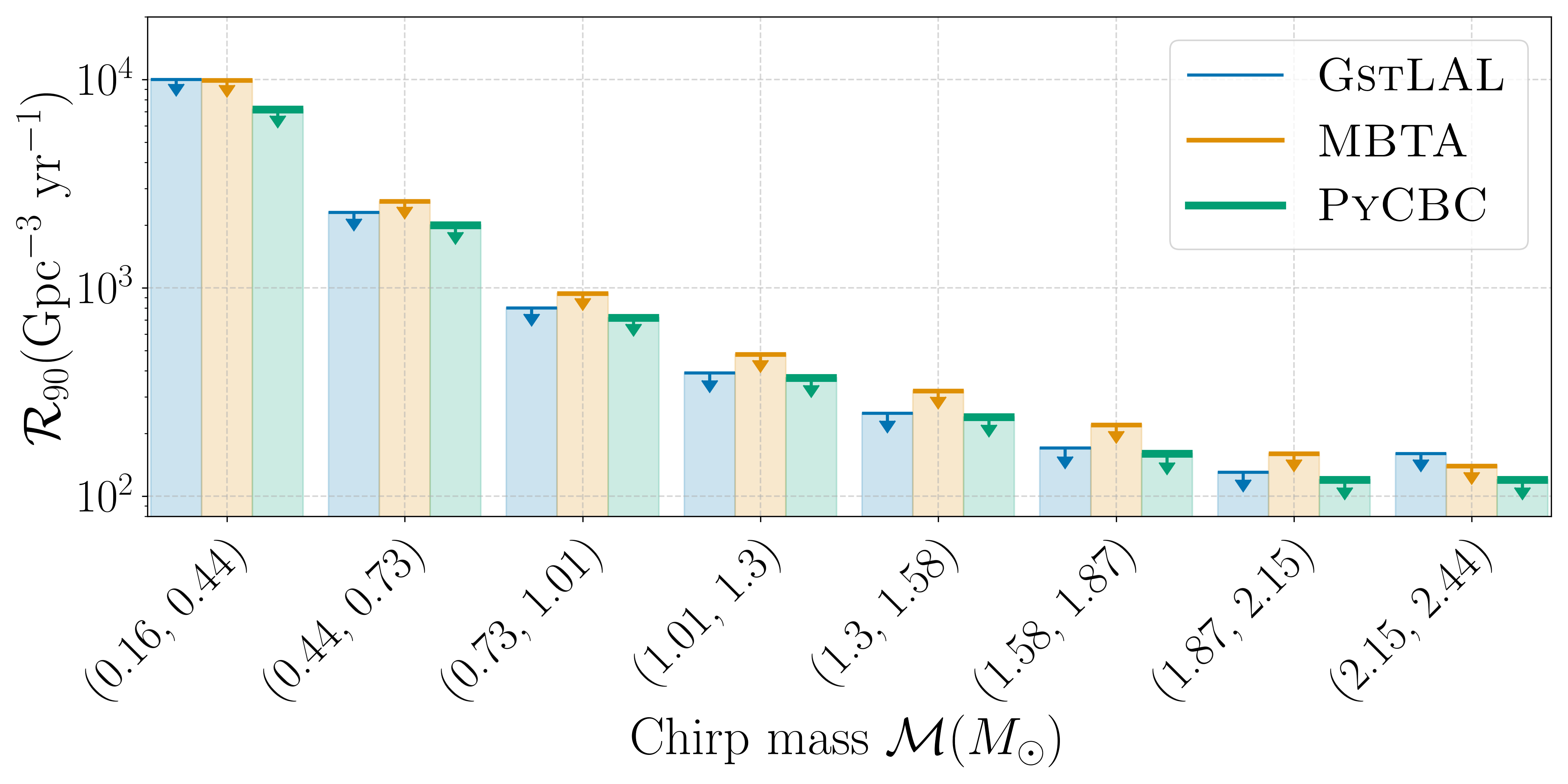}
    \caption{Upper limits on the merger rate of \ac{SSM} \ac{BBH} at the 90\% confidence level, $\mathcal{R}^{\rm O4a}_{90}~(\rm Gpc^{-3}\,yr^{-1})$, at the end of \ac{O4a} as a function of the chirp mass \Mc for search pipelines \GSTLAL (\textit{blue}), \MBTA (\textit{yellow}) and \PYCBC (\textit{green}). The upper limits are computed following Equation~\ref{eq:rate_limit} using the estimated $\langle VT \rangle$ within chirp mass bins specified along the x-axis. The arrows point in the direction of possible merger rates informed by $\langle VT\rangle_{\rm O4a}$ for each search.}
    \label{fig:Rate_mc}
\end{figure}

Figure~\ref{fig:Rate_mc} shows the upper limit on the merger rate at the 90\% confidence level for \GSTLAL, \MBTA, and \PYCBC as a function of the chirp mass \Mc.
Since $\mathcal{R}_{90,i}$ is derived through a rescaling of the $\langle VT \rangle_i$, we find that using data from \ac{O4a} alone yields only marginal improvement relative to \ac{O3}.
We find the most constraining limit on $\mathcal{R}^{\rm O4a}_{90}$ to be \MostConstrainedRateMcGSTLAL, \MostConstrainedRateMcMBTA,~and \MostConstrainedRateMcPYCBC for \GSTLAL, \MBTA, and \PYCBC respectively.

\begin{figure}[t]
    \centering
        \includegraphics[width=\columnwidth]{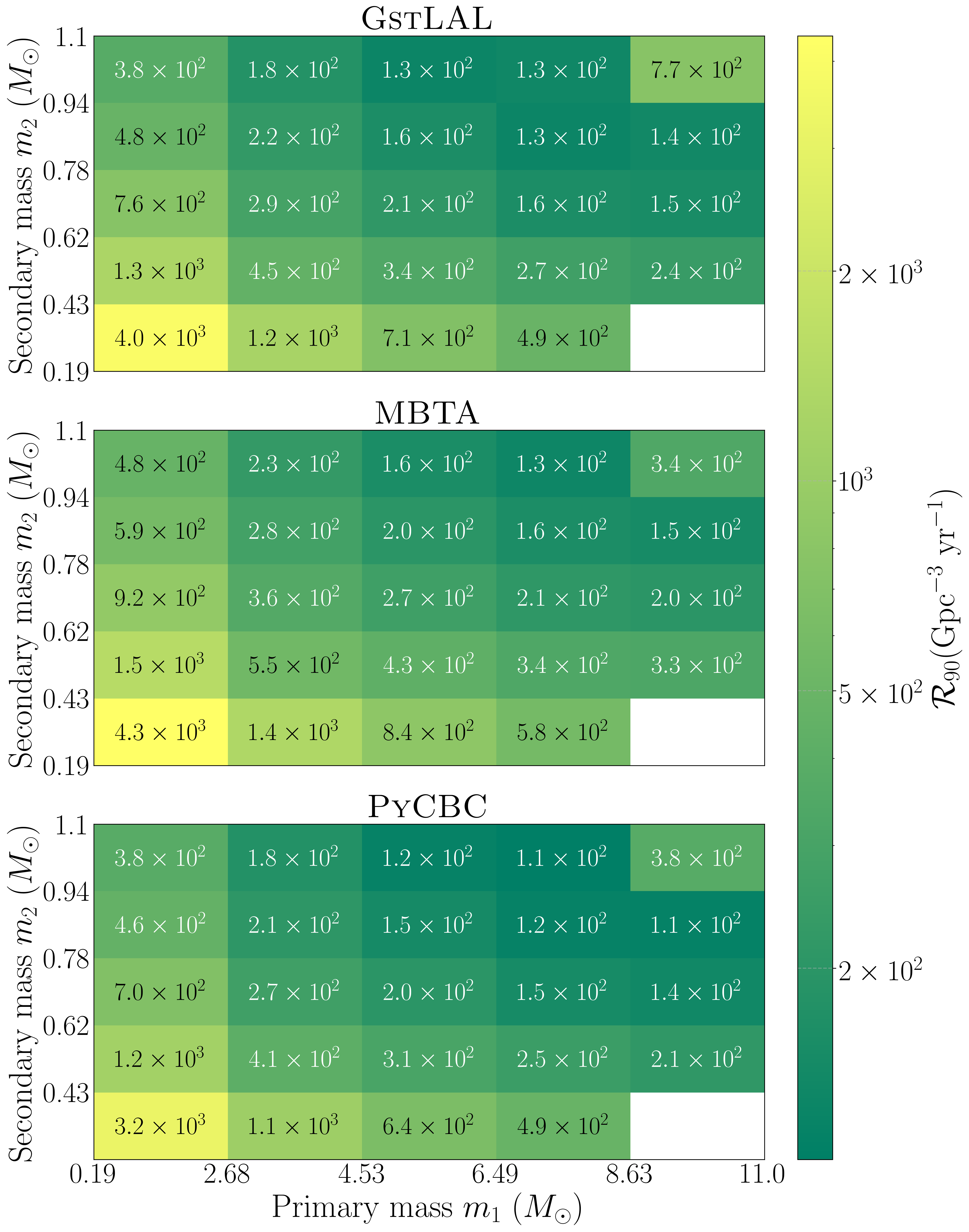}
    \caption{Upper limits on the merger rate of \ac{SSM} \ac{BBH} at the 90\% confidence interval, $\mathcal{R}_{90}~(\rm{Gpc^{-3}\,yr^{-1}})$ as a function of the component masses, (\massone, \masstwo), computed for searches \GSTLAL, \MBTA and \PYCBC. The color of each cell corresponds to the $\mathcal{R}^{\rm O4a}_{90}$ calculated using Equation~\ref{eq:rate_limit} with $\langle VT \rangle_{\rm O4a}$ shown in Figure~\ref{fig:VT_m1m2}.}
    \label{fig:Rate_m1m2}
\end{figure}

We also report upper limits on $\mathcal{R}^{\rm O4a}_{90}$ as a function of (\massone, \masstwo) as shown in Figure~\ref{fig:Rate_m1m2}.
Following the binning in the \masscomponent parameter space described in Section~\ref{sec:bbh_vt}, $\mathcal{R}^{\rm O4a}_{90,i}$ are computed using Equation~(\ref{eq:rate_limit}) in each bin.
We find the most constraining limit on $\mathcal{R}^{\rm O4a}_{90}$ to be \MostConstrainedRateCompmassGSTLAL, \MostConstrainedRateCompmassMBTA,~and \MostConstrainedRateCompmassPYCBC for \GSTLAL, \MBTA, and \PYCBC respectively.

We include the $1\sigma$ uncertainties propagated from the uncertainty in $\langle VT \rangle$ to derive a conservative upper limit.
Propagating the $\langle VT\rangle_{\rm O4a}$ uncertainty (Equation 2) expectedly yields $1\sigma$ statistical errors on the \ac{BBH} merger rate upper limits between $2\%$ and $8\%$.
Note that systematic uncertainties on $\langle VT\rangle_{\rm O4a}$ result in equivalent uncertainties on the rate upper limits.

Similar to the estimation of the merger rate for \ac{BNS} assuming a simple, fixed population in~\cite{LIGOScientific:2025pvj}, we estimate the merger rate for \ac{SSM} \acp{BNS}.
Following the \textsc{SIMPLE UNIFORM BNS} model with modifications to the range of component masses, we assume a uniform mass distribution between $0.5\,\Msun$ and $2.5\,\Msun$ for \massone, and a uniform mass distribution between $0.5\,\Msun$ and $1\,\Msun$ for \masstwo with \masstwo $\leq$ \massone, isotropically distributed spins with uniform spin magnitudes below $0.4$, and a merger rate uniform in comoving volume up to $\redshift_{\rm max}\,=\,0.2$. 
Under this fiducial model with zero \ac{SSM}-\ac{BNS} detections, we use Equation~\ref{eq:rate_limit} to compute the upper limit at the loudest event \ac{FAR}.
The upper limit at the 90\% confidence level is \BNSRateGSTLAL~${\rm Gpc^{-3}\,yr^{-1}}$, \BNSRateMBTA~${\rm Gpc^{-3}\,yr^{-1}}$ and \BNSRatePYCBC~${\rm Gpc^{-3}\,yr^{-1}}$, evaluated at a \ac{FAR} threshold of \GSTLALLoudestTriggerFAR, \MBTALoudestTriggerFAR~and \PYCBCLoudestTriggerFAR~for \GSTLAL, \MBTA and \PYCBC respectively.
Although the merger rate we estimate is broadly consistent with the reported rate for the \textsc{SIMPLE UNIFORM BNS} model in~\cite{LIGOScientific:2025pvj}, the two are not directly comparable due to the differences in assumptions about the underlying \ac{BNS} populations.

%% file: dm_constraints.tex
\section{Constraining Dark Matter}\label{sec:dmconstraints}

The upper limits on merger rates derived in Section~\ref{sec:rates} can be cast into constraints on any model that predicts an observable number of binary mergers within the considered sub-solar mass search parameter space.
In this work we consider two possible sources of \ac{BBH} mergers, \acp{PBH} in Section~\ref{sec:dmconstraints:PBH} and \acp{DBH} in Section~\ref{sec:dmconstraints:DBH}.
For each model, we present the constraints in terms of the maximum fraction of dark matter allowed by our observations that can be composed of \acp{BH} below a solar mass.

\subsection{Primordial Black Holes}\label{sec:dmconstraints:PBH}

In the literature, numerous \ac{PBH} models have been proposed~\citep{Green:2020jor,Carr:2023tpt,LISACosmologyWorkingGroup:2023njw}, each predicting a different \ac{PBH} abundance characterized by the fraction of the dark matter cosmological density composed of \acp{PBH}, defined as $f_\mathrm{PBH} = \rho_\mathrm{PBH}/\rho_\mathrm{DM}$,
where $\rho_\mathrm{PBH}$ and $\rho_\mathrm{DM}$ are the present-day mass densities of \acp{PBH} and dark matter, respectively.
Constraining $f_\mathrm{PBH}$ robustly using \ac{GW} observations is challenging because, for a given $f_\mathrm{PBH}$, the predicted binary merger rates at the present epoch vary significantly across models~\citep{Nakamura:1997sm,Ali-Haimoud:2017rtz,Raidal:2018bbj,Raidal:2017mfl,Bird:2016dcv,Clesse:2016vqa,Clesse:2020ghq}.
These variations primarily arise from differences in the initial mass distributions and clustering properties of \acp{PBH}~\citep{Chen:2018czv,Clesse:2016vqa,Ballesteros:2018swv,Young:2019gfc,Siles:2024yym}.
Moreover, the merger rate predictions remain highly uncertain as they are strongly influenced by complex phenomena such as many-body interactions and non-linear gravitational effects, which are difficult to model accurately~\citep{Raidal:2018bbj,Vaskonen:2019jpv,Jedamzik:2020ypm,Jedamzik:2020omx}.

\ac{PBH} binaries are typically considered to form in two scenarios, early binaries (EB) and late binaries (LB).
Early binaries form shortly after \ac{PBH} formation as pairs of \acp{PBH} decouple from the Universe's expansion and form eccentric binaries through tidal interactions with nearby \acp{PBH}~\citep{Nakamura:1997sm}.
Consequently, merger rates of early binaries are strongly influenced by initial \ac{PBH} clustering and mass distribution.
In contrast, late binaries form in dense halos through dynamical interactions, with merger rates shaped by local density and halo properties~\citep{Siles:2024yym}.

We model the merger rate of early binaries as in~\citet{LIGOScientific:2022hai}, using the analytical approximations developed in~\citet{Hutsi:2020sol,Vaskonen:2019jpv,Chen:2018czv,Ali-Haimoud:2017rtz} and validated by $N$-body simulations in~\citet{Raidal:2018bbj}
\footnote{The merger-rate expression in Eq.~\eqref{eq:Rate_EB}, while widely used in the literature, is validated using $N$-body simulations in a limited region of parameter space. 
In particular, the simulations of~\citet{Raidal:2018bbj} consider $\sim\mathcal{O}(10^2)$ equal-mass $\sim 30\,\Msun$ PBHs with $f_\mathrm{PBH} \lesssim 0.1$, which differs significantly from the lower masses and larger PBH abundances explored here.}:

\begin{align}
	\frac{\mathrm{d}^2\mathcal{R}^\mathrm{PBH}_\mathrm{EB}}{\mathrm{d} \ln m_A \mathrm{d} \ln m_B} =& 1.6 \times 10^6 \, \mathrm{Gpc}^{-3} \, \mathrm{yr}^{-1} \times f_\mathrm{sup} f_\mathrm{PBH}^{53/37} f(\ln m_A)  \nonumber \\
	& \times f(\ln m_B) \left( \frac{m_A+m_B}{\Msun} \right)^{-32/37} \left[ \frac{m_A m_B}{(m_A+m_B)^2}\right]^{-34/37},  \label{eq:Rate_EB}
\end{align}

\noindent \noindent where $m_A$ and $m_B$ denote the component masses of the \ac{PBH} binary, without imposing any ordering between them, in contrast to the convention $\massone \geq \masstwo$ used elsewhere in this work.
 $f(\ln m)$ is the normalized \ac{PBH} mass function (i.e., the probability density over $\ln m$), and $f_\mathrm{sup}$ is a suppression factor that accounts for binary disruption due to gravitational interactions with other \acp{PBH} or surrounding matter.
To remain consistent with~\citet{LIGOScientific:2022hai}, we adopt $f_\mathrm{sup} = 2.3 \times 10^{-3} f_\mathrm{PBH}^{-0.65}$ throughout this work, representative of a high-suppression scenario~\citep{Hutsi:2020sol}, and thus leading to conservative constraints.
Since this high-suppression regime becomes relevant only at large \ac{PBH} abundances~\citep{Hutsi:2020sol}, whereas microlensing surveys already restrict $f_\mathrm{PBH} \lesssim 0.1$ in this mass range~\citep[e.g.,][]{Carr:2023tpt,Green:2020jor}, our choice is likely over-conservative.

The merger rate of late binaries is significantly more challenging to model, as this channel involves complex dynamical interactions. Following~\citet{Clesse:2020ghq,Phukon:2021cus}, we assume

\begin{align}
	\frac{\mathrm{d}^2\mathcal{R}^\mathrm{PBH}_\mathrm{LB}}{\mathrm{d} \ln m_A \mathrm{d} \ln m_B} = R_\mathrm{clust} f_\mathrm{PBH} f(\ln m_A) f(\ln m_B) \frac{(m_A + m_B)^{10/7}}{(m_A m_B)^{5/7}},  \label{eq:Rate_LB}
\end{align}

\noindent where $R_\mathrm{clust}$ is a scaling factor that depends on the \ac{PBH} clustering properties, including their velocity distribution.
While there is general agreement on the mass dependence of the late binaries' merger rate~\citep[e.g.,][]{Bird:2016dcv}, the overall amplitude remains highly uncertain due to its strong sensitivity to the underlying clustering scenario.
As in~\citet{Clesse:2020ghq,Phukon:2021cus}, we assume $R_\mathrm{clust} = 420 \, \mathrm{Gpc}^{-3} \, \mathrm{yr}^{-1}$, which lies toward the upper end of the plausible range~\citep{Phukon:2021cus} and corresponds to an optimistic, though physically motivated, clustering scenario.
Since the predicted rate scales linearly with $R_\mathrm{clust}$, this choice leads to stronger constraints than a low-clustering assumption.

We use the theoretical merger rate models of Equations~(\ref{eq:Rate_EB},~\ref{eq:Rate_LB}) to set upper limits on $f_\mathrm{PBH}$, assuming a point \ac{PBH} mass distribution, i.e., $f(\ln m) = \delta(\ln m - \ln m_\mathrm{PBH})$.
The limits are obtained by identifying the value of $f_\mathrm{PBH}$ for which the theoretical rates match the 90\% confidence level upper bounds on merger rates derived using Equation~\eqref{eq:rate_limit}.

\begin{figure}
    \centering
    \includegraphics[width=1\columnwidth]{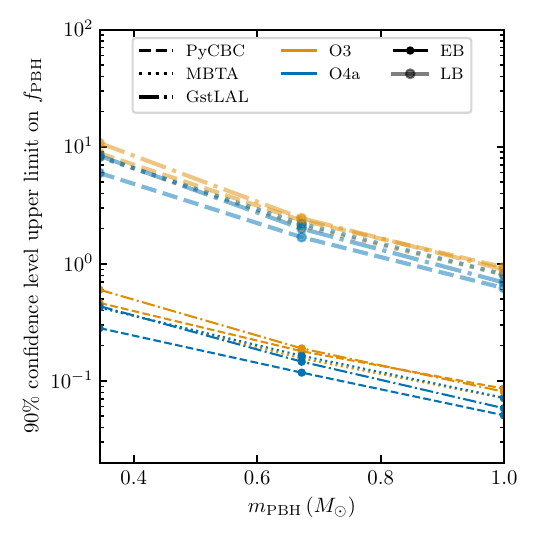}
    \caption{
    Upper limits on the \ac{PBH} dark matter fraction $f_{\text{PBH}}$ as a function of \ac{PBH} mass $m_{\text{PBH}}$.
    Constraints are derived using the merger rate models given by Equation~\eqref{eq:Rate_EB} for early binaries (EB), and Equation~\eqref{eq:Rate_LB} for late binaries (LB), assuming a a point mass function.
    Results are shown at the 90\% confidence level using the sensitive hypervolumes $\langle VT \rangle$ calculated for \GSTLAL, \MBTA, and \PYCBC during \ac{O3} and \ac{O4a}.
    }
    \label{fig:fPBH_O4a_withLB}
\end{figure}

Figure~\ref{fig:fPBH_O4a_withLB} shows the resulting constraints for both early binaries and late binaries, using the $\langle VT \rangle$ of the \GSTLAL, \MBTA, and \PYCBC pipelines in \ac{O3} and \ac{O4a}.
We find close agreement among the different pipelines. 
The constraints from \ac{O3} and \ac{O4a} are of similar magnitude, reflecting the comparable $\langle VT \rangle$s in the two observing runs, though \ac{O4a} consistently provides slightly stronger limits.
These direct search constraints are complementary to the bounds derived from the non-observation of a stochastic \ac{GW} background from inspiralling \ac{SSM} \ac{PBH} binaries at high redshift~\citep{Boybeyi:2024mhp,LIGOScientific:2025kry}.



While the constraints presented above assume a point \ac{PBH} mass function, this is not realistic in most early-formation scenarios~\citep{Carr:2023tpt}.
\acp{PBH} are generally expected to form with broad and non-trivial mass distributions~\citep{LISACosmologyWorkingGroup:2023njw}.
To remain agnostic about the specific shape of the mass function, we instead constrain the quantity

\begin{align}
	F_\mathrm{PBH} & = \frac{f_\mathrm{sup}}{2.3\times 10^{-3}} f_\mathrm{PBH}^{53/37} f(\ln m_A) f(\ln m_B) \nonumber \\
	& = f_\mathrm{PBH}^{0.78} f(\ln m_A) f(\ln m_B) ,  \label{eq:FPBH_def}
\end{align}

\noindent which encapsulates the dependence of the merger rate of early binaries on both the \ac{PBH} abundance $f_\mathrm{PBH}$, and the underlying mass distribution $f(\ln m)$.
The definition of $F_\mathrm{PBH}$ explicitly incorporates the assumption $f_\mathrm{sup} = 2.3 \times 10^{-3} f_\mathrm{PBH}^{-0.65}$ adopted throughout this work.

Limits on $F_\mathrm{PBH}$ are determined by finding the value for which the theoretical merger rate for early binaries matches the 90\% confidence level upper bounds derived from Equation~\eqref{eq:rate_limit}.
We constrain $F_\mathrm{PBH}$ only for early binaries, since its definition in Equation~\eqref{eq:FPBH_def} is specific to this scenario, and merger rate models for late binaries are more uncertain and less constraining.
We adopt the binning in (\massone, \masstwo) from Section~\ref{sec:rates}, and write the integral in Equation~\eqref{eq:Rate_EB} within each bin as

\begin{align}
	\mathcal{R}^\mathrm{PBH}_\mathrm{EB} = & 3.68 \times 10^3 \, \mathrm{Gpc}^{-3} \, \mathrm{yr}^{-1} \overline{F}_\mathrm{PBH} \times  \left( \frac{\overline{m}_1 + \overline{m}_2}{\Msun} \right)^{-32/37} \nonumber \\
	& \times \left[ \frac{\overline{m}_1 \overline{m}_2}{(\overline{m}_1 + \overline{m}_2)^2}\right]^{-34/37}  \ln\left( \frac{m_1^\mathrm{max}}{m_1^\mathrm{min}}\right) \ln\left( \frac{m_2^\mathrm{max}}{m_2^\mathrm{min}}\right) ,  \label{eq:Rate_EB_FPBH}
\end{align}

\noindent where $\masscomponent^\mathrm{min}$ and $\masscomponent^\mathrm{max}$ denote the bin edges for the component masses, $\overline{m}_{1,2} = (\masscomponent^\mathrm{min} + \masscomponent^\mathrm{max})/2$ is the central value, and $\overline{F}_\mathrm{PBH}$ denotes the effective bin-averaged value of $F_\mathrm{PBH}$, defined implicitly such that Eq.~\eqref{eq:Rate_EB_FPBH} reproduces the exact integral of the merger rate over the bin.

\begin{figure}
    \centering
    \includegraphics[width=1\columnwidth]{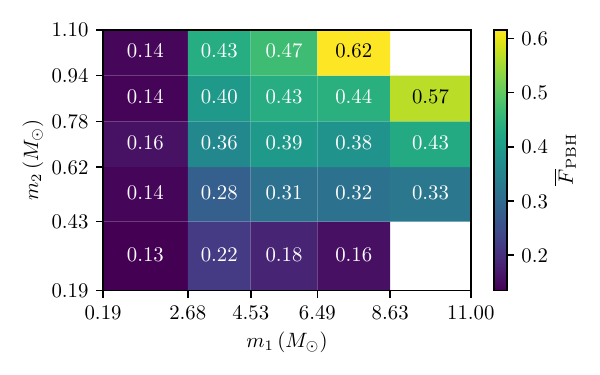}
    \caption{
	Upper limits on $\overline{F}_\mathrm{PBH}$ (Equation~\eqref{eq:FPBH_def}) as a function of the \ac{PBH} binary component masses \massone and \masstwo.
	Constraints are derived using the early binary merger rate model of Equation~\eqref{eq:Rate_EB_FPBH}.
    We show the median 90\% upper limit across \GSTLAL, \MBTA, and \PYCBC calculated using their sensitive hypervolumes $\langle VT \rangle$ in \ac{O4a}.}
    \label{fig:FPBH_m1m2_Cumulative_EB}
\end{figure}

Figure~\ref{fig:FPBH_m1m2_Cumulative_EB} shows the median 90\% upper limit on $\overline{F}_\mathrm{PBH}$ across all search pipelines, based on the results from \ac{O4a}.
We find that $\overline{F}_\mathrm{PBH}$ is constrained to be less than unity across the entire range of (\massone, \masstwo) explored in this work, though the limits remain relatively weak as they lie close to unity.

\subsection{Dissipative Dark Matter}\label{sec:dmconstraints:DBH}

Dissipative dark matter is a class of models which posit that dark matter may have complex particle interactions similar to those of the Standard Model, including dissipative processes.
Whenever there is a fermion in the model, dark matter has a Chandrasekhar limit analogous to that of the Standard Model, but with a value that depends on the dark matter particle mass(es)~\citep{Mohapatra:1999ih, DAmico:2017lqj, Shandera:2018xkn, Choquette:2018lvq, Chang:2018bgx, Bramante:2023ddr}.
In dissipative scenarios, particle interactions allow energy to be transported away from gravitationally bound systems such as dark matter halos.
These dissipative interactions become more efficient in denser regions, and some dense regions may dissipate energy efficiently enough for gravity to dominate, leading to the formation of ultra-compact objects made of dark matter.
When those are sufficiently massive, above the dark Chandrasekhar limit, they are \aclp{DBH}.
Therefore, \ac{GW} signals from \ac{DBH} binaries provide a way to probe the microphysics of dissipative dark matter even if it is not coupled to the Standard Model beyond gravity, or if the coupling is too weak to be detectable in the near future.

Although the fragmentation scale of the dark gas depends on the chemistry of the dark sector, the dynamical processes that lead to the formation of compact objects are analogous to those in star formation.
While simulation of such small-scale phenomena in novel dark matter scenarios remains challenging, it is reasonable to expect that the initial mass function and binary fraction of \acp{DBH} are qualitatively similar to that of baryonic objects.
For the particular worked example of atomic dark matter~\citep{Ackerman:2008kmp, Kaplan:2011yj, Cyr-Racine:2012tfp}, where cooling and compact object formation has been extensively studied~\citep{Rosenberg:2017qia,Ryan:2021dis, Gurian:2021qhk, Ryan:2021tgw}, there is a close parallel between \ac{DBH} formation and Population III star formation that further supports this statement.
Thus, we use a truncated power-law mass model characterized by a minimum mass $M_\text{min}$, power $-b$, and ratio of maximum mass to minimum mass $r = M_\text{max}/M_\text{min}$, and assume a binary fraction of $f_\text{binary} = 0.26$~\citep{Shandera:2018xkn} in our analysis.
The total number of \acp{DBH} in this population is set by $f_\mathrm{DBH} \equiv \rho_\text{DBH}/\rho_\text{DM}$ where $\rho_\text{DBH}$ and $\rho_\text{DM}$ are the densities of \acp{DBH} and dark matter, respectively.
The rate of \ac{DBH} binaries merging in this population, $\mathcal{R}$, is computed as~\citep{Singh:2020wiq}

\begin{equation}
\begin{split}
\frac{\mathrm{d}\mathcal{R}(\mathcal{M}\mid f_{\mathrm{DBH}}, \theta)}{\mathrm{d}\mathcal{M}}
& = \Bigl[ p(\mathcal{M} \mid t_m, \theta) \times p(t_m \mid \theta) \Bigr]_{t_m = 10\,\mathrm{Gyr}} \\
& \;\;\;\; \times \left(\frac{\rho_{\mathrm{DM}} \times f_{\mathrm{DBH}} \times f_{\mathrm{binary}}}{\langle M \rangle}\right),
\end{split}
\label{eq:peters_formula}
\end{equation}

\noindent where $p(\mathcal{M} \mid t_m, \theta)$ is the distribution of \Mc with a time of merging $t_m$, $p(t_m \mid \theta)$ is the probability distribution of times-to-merger (these two factors evaluated at $t_m = 10\,\mathrm{Gyr}$), $\rho_\text{DM} = 3.3 \times 10^{19}\,\Msun~\text{Gpc}^{-3}$, and $\langle M \rangle$ is the average component mass of the \ac{DBH} binaries.
The merger times are calculated using Peters formula for coalescing binaries~\citep{Peters:1963ux}.

We use a Bayesian analysis to constrain this population of \acp{DBH} by creating \Mc bins and modeling the number of detected binaries in each bin using a Poisson distribution.
With the non-detection of \ac{SSM} events, the likelihood in the $i$--th chirp-mass bin is $P(0|\mathcal{R}_i,\langle VT \rangle_i,\Mc) = e^{-\mathcal{R}_i \langle VT \rangle_i}$, where the rate $\mathcal{R}_i$ is determined by integrating Equation (\ref{eq:peters_formula}) over the width of the $i$--th \Mc bin, and the sensitive hypervolume $VT_i$ is informed by this \ac{SSM} search.
Because we adopt a Bayesian framework with a Poisson likelihood conditioned on zero detections, the sensitive hypervolume must be evaluated at a fixed, confident detection threshold rather than via a loudest-event-statistic method, which would conflate confidence levels with credible intervals.
We therefore adopt a more stringent \ac{FAR} threshold of \FARThreshVTDBH~when estimating the pipeline-specific $\langle VT \rangle_{\rm O4a}$, ensuring an unambiguous null-detection result in every \Mc bin.
We use priors consistent with previous analyses~\citep{LIGOScientific:2021job,LIGOScientific:2022hai}: uniform priors on $M_\text{min}$ over the interval $[10^{-3}\,\Msun, 3.1\,\Msun ]$, $b$ over the interval $[-1,2]$, and $f_\text{DBH}$ over the interval $[10^{-10},1]$, and a log-uniform prior on $r$ over the interval $[2, 1000]$.
Bayes' law then states the log-posterior is

\begin{equation}
    \ln p(f_\text{DBH}, M_\text{min}, r,b) =  -\sum_i \mathcal{R}_i(f_\text{DBH},M_\text{min},r,b)\langle VT \rangle_i  -\ln r + C,
\end{equation}

\noindent where $C$ is the normalization constant.

\begin{figure}
    \centering
    \includegraphics[width=1\columnwidth]{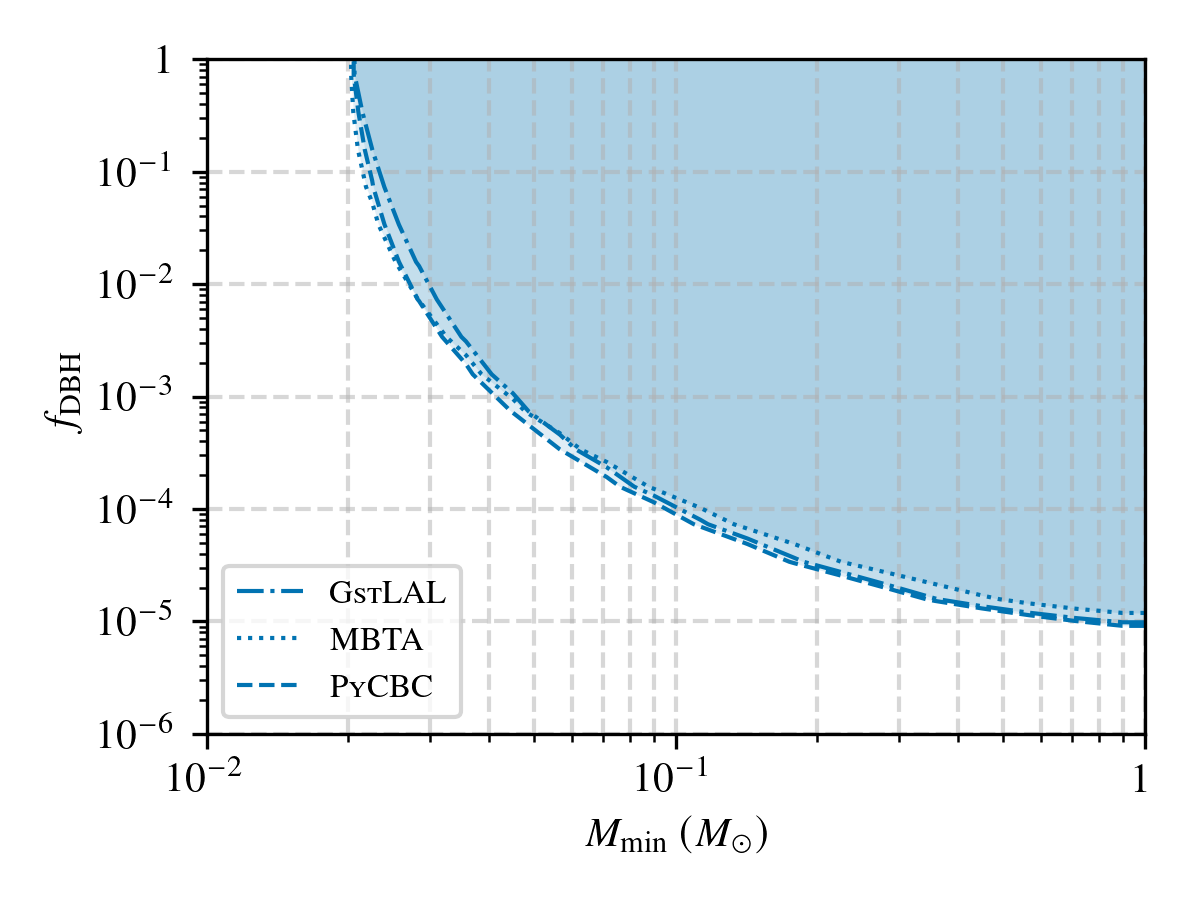}
    \caption{
        Contours in the $f_{\rm DBH}$--$M_{\min}$ plane for the atomic dark matter model after marginalizing over the ratio of maximum to minimum mass, and the slope of the black hole population.
        Using the two-dimensional PDF $p(f_{\rm DBH}, M_{\min})$, we calculate 90\% of the posterior mass to be outside of the shaded region.
        Constraints shown are derived from $\langle VT\rangle_{\rm O4a}$ for the \GSTLAL, \MBTA, and \PYCBC pipelines evaluated with a \ac{FAR} threshold of \FARThreshVTDBH.
        We note that the apparent ordering of the pipeline constraints at small $M_{\min}$ is due to finite posterior sampling rather than a physical difference in sensitivity.}
    \label{fig:atomic_dark_matter_f_vs_mmin}
\end{figure}

Previous analyses evaluated the posterior distribution using a grid-based method~\citep{LIGOScientific:2021job,LIGOScientific:2022hai}, while we use the MCMC sampler \textsc{emcee}~\citep{Foreman-Mackey:2012any} to explore the parameter space for faster convergence in this analysis.
We marginalize over $r$ and $b$ to constrain $f_\text{DBH}$ and $M_\text{min}$.
Figure~\ref{fig:atomic_dark_matter_f_vs_mmin} shows the two-dimensional PDF of $f_\text{DBH}$ and $M_\text{min}$ derived from $\langle VT \rangle_{\rm O4a}$ for \GSTLAL, \MBTA, and \PYCBC individually.
Given no detections, the resulting posterior distribution can be interpreted as upper limits on $f_\text{DBH}$.

The exclusion region in the $f_{\rm DBH}$--$M_{\min}$ space extends as low as $f_{\rm DBH} = \fconstraintOfoura$ at $M_{\min} = 1\,\Msun$, where the range arises from the differing sensitivities of the search pipelines.
This limit on $f_\text{DBH}$ is consistent with the dissipative dark matter scenario as the formation channel requires efficient cooling, which only occurs in regions dense in dark matter.
Since we restrict our prior on $r \leq 1000$, a non-detection provides no information below \mminlowerbound because the searches have minimal support for mass distributions with lower minimum masses.
Additionally, we exclude limits where $M_\text{min} > 1\,\Msun$ as an \ac{SSM} detection necessitates the minimum mass to be smaller than $1\,\Msun$.
The range of dark fermion masses $m_\chi$ probed by this search is set by $M_{\rm min}$ via dark Chandrasekhar mass, $M_\text{Chandra}^\text{Dark} \propto M_\text{Planck}^3/(\mu^\text{Dark}m_\chi)^2 \leq M_\text{min},$ where $\mu^\text{Dark} m_\chi$ is the mass per degenerate fermion~\citep{Chandrasekhar:1931ih}.
We derive a bound on the ratio of $m_\chi$ to the mass of the Standard Model proton $m_p$ by taking the ratio of the stellar Chandrasekhar mass to the dark Chandrasekhar mass as

\begin{equation}
    \mu^\text{Dark}\frac{m_\chi}{m_p} = 2\sqrt{\frac{1.4\,\Msun}{M_\text{Chandra}^\text{Dark}}} \leq 2 \sqrt{\frac{1.4\,\Msun}{M_\text{min}}},
\end{equation}

\noindent where we have set the stellar Chandrasekhar mass to $1.4\,\Msun$, and corresponding $\mu^\text{Stellar}$ to $2$.

Previous searches~\citep{LIGOScientific:2022hai} used the Standard Model value of $\mu^\text{Dark}=2$, but we report the bound leaving $\mu^\text{Dark}$ unspecified to cover a broader range of dissipative dark matter models.
The search thus probed $\mu^\text{Dark}m_\chi/m_p$ in the range $ \darkprotonlower < \mu^\text{Dark}m_\chi/m_p < \darkprotonupper$ where the range arises from the differing sensitivities of the search pipelines.
A future detection would be able to put a direct bound on the microphysics of dissipative dark matter.

%% file: conclusion.tex
\section{Conclusion}\label{sec:conclusion}

We present results from \ac{SSM} searches for compact objects in data from the first part of the fourth LIGO--Virgo--KAGRA observing run~\acs{O4a}.
No significant candidate, other than the previously published candidate GW230529\_181500~\citep{LIGOScientific:2024elc}, is reported by the three participating searches.
Although all searches detect GW230529\_181500, it is excluded as a possible \ac{SSM} detection because the component masses of the source binary are reliably known to be greater than a solar mass.
Given the null detection result, we report upper limits on the merger rates for \ac{SSM} \acp{BBH} with masses greater than $0.2\,\Msun$ using the estimated search sensitivity.

Using the inferred sensitive hypervolume and limits on the merger rate of \ac{SSM} \acp{BBH}, we update the constraints on two dark matter models: \acp{PBH}, and \acp{DBH} that form in a dissipative dark matter model.
While both models can populate the sub-solar and super-solar regime, we turn to the \ac{SSM} parameter space for clear evidence of such objects because no known astrophysical channel produces \ac{SSM} black holes.
\ac{O4a} data alone provides only marginal improvement over previous limits~\citep{LIGOScientific:2022hai}.

The sensitive hypervolume is subject to both statistical and systematic uncertainties that directly influence the resulting merger rate upper limits and constraints on dark matter models.
Statistical uncertainty from the finite injection sampling is compounded by effects like calibration errors, and waveform systematics.
We adopt a more conservative, binned approach to report the sensitivity of searches ensuring the limits remain robust against sampling fluctuations.

In the \ac{PBH} scenario, we set informative constraints on the fraction of dark matter in \acp{PBH} $f_\mathrm{PBH}$ in late-forming binaries for the first time, with $f_\mathrm{PBH} < 1$ at the 90\% C.L. for $m_\mathrm{PBH} \gtrsim \LBmPBHfPBHconstrainedOfoura$ using the $\langle VT \rangle$ from \ac{O4a}.
For early-forming \ac{PBH} binaries, this search improves upon previous bounds under a conservative model that accounts for binary disruption, with $f_\mathrm{PBH}$ upper limits between $\EBfPBHMinMPBHOfoura$ and $\EBfPBHMaxMPBHOfoura$ over the covered range of \ac{PBH} masses.
For the dissipative dark matter scenario, we find that \ac{SSM} \acp{DBH} can only form $\fconstraintOfoura$ of the total dark matter density in the Universe.

For the first time, we report the sensitivity of \ac{LVK} \ac{SSM} searches to binaries containing low-mass neutron stars.
While~\cite{2025ApJ...984...61K,Kacanja:2026byy} have performed searches for ultra low-mass \ac{NS} in data from \ac{O3} and \ac{O4a} that include tides in the template bank construction itself, we determine the sensitivity of \ac{SSM} searches designed for \acp{BBH} towards binaries with sub-solar \acp{NS} where we only include tides in the simulated signals.
We find that the typical sensitive hypervolume is $\langle VT \rangle$ \BNSVTorderofmagnitude~for populations of \acp{BNS} with at least one \ac{SSM} neutron star.
These searches can also provide insight into matter at high densities given the overlap of the neutron star and \ac{SSM} parameter space, serving as motivation for low-latency \ac{SSM} searches~\citep{Allene:2025saz, Hanna:2024tom} which started to release alerts in early 2025.

Recently,~\cite{Kacanja:2026byy} reported results from an \ac{SSM} search that includes tidal effects for low mass neutron stars covering binaries with $0.1\,\Msun \leq \massone (\masstwo) \leq 2\,\Msun (1\,\Msun)$, and aligned spins $-0.05 \leq \spincomponent \leq 0.05$.
They report no significant candidate from this search, and find consistent limits on the merger rate for overlapping regions of the parameter space.

The LIGO detectors resumed observation with improved sensitivity, and were joined by the Virgo detector in the \ac{O4b}, and by the KAGRA detector in the \ac{O4c}.
With increased observing time as we include data from the later parts of the fourth observing run, we expect the search sensitivity to increase by a factor of 2 to 3 leading to improved constraints on the merger rate of \ac{SSM} \acp{BBH}.
These advances will provide more stringent limits on the fraction of dark matter in \ac{SSM} compact objects, or perhaps even detect a gravitational wave signal originating from the merger of \ac{SSM} compact objects.
Such a detection would have deep implications for fundamental physics.

%% file: data.tex
All strain data analysed as part of this publication are publicly available through \ac{GWOSC}.
The details of this data release are described in detail in~\citet{LIGOScientific:2025snk}.
Associated data release includes trigger files for candidates reported in Table~\ref{tab:triggers}, \ac{BBH} sensitive volume-time, and injection results used in the calculation of sensitive volume-time reported in this paper~\citep{ZenodoO4aSSM}.

%% file: LVKack.tex
This material is based upon work supported by NSF's LIGO Laboratory, which is a
major facility fully funded by the National Science Foundation.
The authors also gratefully acknowledge the support of
the Science and Technology Facilities Council (STFC) of the
United Kingdom, the Max-Planck-Society (MPS), and the State of
Niedersachsen/Germany for support of the construction of Advanced LIGO 
and construction and operation of the GEO\,600 detector. 
Additional support for Advanced LIGO was provided by the Australian Research Council.
The authors gratefully acknowledge the Italian Istituto Nazionale di Fisica Nucleare (INFN),  
the French Centre National de la Recherche Scientifique (CNRS) and
the Netherlands Organization for Scientific Research (NWO)
for the construction and operation of the Virgo detector
and the creation and support  of the EGO consortium. 
The authors also gratefully acknowledge research support from these agencies as well as by 
the Council of Scientific and Industrial Research of India, 
the Department of Science and Technology, India,
the Science \& Engineering Research Board (SERB), India,
the Ministry of Human Resource Development, India,
the Spanish Agencia Estatal de Investigaci\'on (AEI),
the Spanish Ministerio de Ciencia, Innovaci\'on y Universidades,
the European Union NextGenerationEU/PRTR (PRTR-C17.I1),
the ICSC - CentroNazionale di Ricerca in High Performance Computing, Big Data
and Quantum Computing, funded by the European Union NextGenerationEU,
the Comunitat Auton\`oma de les Illes Balears through the Conselleria d'Educaci\'o i Universitats,
the Conselleria d'Innovaci\'o, Universitats, Ci\`encia i Societat Digital de la Generalitat Valenciana and
the CERCA Programme Generalitat de Catalunya, Spain,
the Polish National Agency for Academic Exchange,
the National Science Centre of Poland and the European Union - European Regional
Development Fund;
the Foundation for Polish Science (FNP),
the Polish Ministry of Science and Higher Education,
the Swiss National Science Foundation (SNSF),
the Russian Science Foundation,
the European Commission,
the European Social Funds (ESF),
the European Regional Development Funds (ERDF),
the Royal Society, 
the Scottish Funding Council, 
the Scottish Universities Physics Alliance, 
the Hungarian Scientific Research Fund (OTKA),
the French Lyon Institute of Origins (LIO),
the Belgian Fonds de la Recherche Scientifique (FRS-FNRS), 
Actions de Recherche Concert\'ees (ARC) and
Fonds Wetenschappelijk Onderzoek - Vlaanderen (FWO), Belgium,
the Paris \^{I}le-de-France Region, 
the National Research, Development and Innovation Office of Hungary (NKFIH), 
the National Research Foundation of Korea,
the Natural Sciences and Engineering Research Council of Canada (NSERC),
the Canadian Foundation for Innovation (CFI),
the Brazilian Ministry of Science, Technology, and Innovations,
the International Center for Theoretical Physics South American Institute for Fundamental Research (ICTP-SAIFR), 
the Research Grants Council of Hong Kong,
the National Natural Science Foundation of China (NSFC),
the Israel Science Foundation (ISF),
the US-Israel Binational Science Fund (BSF),
the Leverhulme Trust, 
the Research Corporation,
the National Science and Technology Council (NSTC), Taiwan,
the United States Department of Energy,
and
the Kavli Foundation.
The authors gratefully acknowledge the support of the NSF, STFC, INFN and CNRS for provision of computational resources.

This work was supported by MEXT,
the JSPS Leading-edge Research Infrastructure Program,
JSPS Grant-in-Aid for Specially Promoted Research 26000005,
JSPS Grant-in-Aid for Scientific Research on Innovative Areas 2402: 24103006,
24103005, and 2905: JP17H06358, JP17H06361 and JP17H06364,
JSPS Core-to-Core Program A.\ Advanced Research Networks,
JSPS Grants-in-Aid for Scientific Research (S) 17H06133 and 20H05639,
JSPS Grant-in-Aid for Transformative Research Areas (A) 20A203: JP20H05854,
the joint research program of the Institute for Cosmic Ray Research,
University of Tokyo,
the National Research Foundation (NRF),
the Computing Infrastructure Project of the Global Science experimental Data hub
Center (GSDC) at KISTI,
the Korea Astronomy and Space Science Institute (KASI),
the Ministry of Science and ICT (MSIT) in Korea,
Academia Sinica (AS),
the AS Grid Center (ASGC) and the National Science and Technology Council (NSTC)
in Taiwan under grants including the Science Vanguard Research Program,
the Advanced Technology Center (ATC) of NAOJ, 
the Mechanical Engineering Center of KEK
and Vietnam National Foundation for Science and Technology Development 
(NAFOSTED) 103.01-2025.147.

Additional acknowledgements for support of individual authors may be found in the following document: \\
\texttt{https://dcc.ligo.org/LIGO-M2300033/public}.
For the purpose of open access, the authors have applied a Creative Commons Attribution (CC BY)
license to any Author Accepted Manuscript version arising.
We request that citations to this article use 'A. G. Abac {\it et al.} (LIGO-Virgo-KAGRA Collaboration), ...' or similar phrasing, depending on journal convention.


%% file: software_ack.tex
Calibration of the \ac{LIGO} strain data was performed with a \GSTLAL{}-based
calibration software pipeline~\citep{Viets:2017yvy}.
Data-quality products and event-validation results were computed using the
\textsc{DMT}{}~\citep{DMTdocumentation}, \textsc{DQR}{}~\citep{DQRdocumentation},
\textsc{DQSEGDB}{}~\citep{Fisher:2020pnr}, \textsc{gwdetchar}{}~\citep{gwdetchar-software},
\textsc{hveto}{}~\citep{Smith:2011an}, \textsc{iDQ}{}~\citep{Essick:2020qpo},
\textsc{Omicron}{}~\citep{Robinet:2020lbf}, \textsc{PEMCHECK}~\citep{Helmling-Cornell:2023wqe} and
\textsc{PythonVirgoTools}{}~\citep{pythonvirgotools} software packages and contributing
software tools.
Analyses in this catalog relied upon the \LALSUITE{} software
library~\citep{lalsuite, Wette:2020air}.
The detection of the signal candidates and subsequent
significance evaluations in this paper were performed with the
\GSTLAL{}-based inspiral software
pipeline~\citep{Messick:2016aqy,Sachdev:2019vvd,Hanna:2019ezx,Cannon:2020qnf},
with the \MBTA{} pipeline~\citep{Adams:2015ulm,Aubin:2020goo}, and with the
\PYCBC{}~\citep{Usman:2015kfa,Nitz:2017svb,Davies:2020tsx}.
Sensitive Hypervolume calculations were performed using \GWPOPULATION~\citep{Talbot2025}.
Plots were prepared with \MATPLOTLIB{}~\citep{Hunter:2007ouj},
\SEABORN{}~\citep{Waskom:2021psk} and \GWPY{}~\citep{gwpy-software}.
\NUMPY{}~\citep{Harris:2020xlr}, \SCIPY{}~\citep{Virtanen:2019joe} and \PANDAS{}~\citep{the_pandas_development_team_2024_10537285} were used
in the preparation of the manuscript.

%% file: VT_methods.tex
\section{Methods to determine the Search Sensitivity}\label{appendix:vt}

To quantify the sensitivity towards a population with model parameters $\mathbf{\Lambda}$ that determine the population distribution of masses and spins, we determine the fraction of detectable mergers $\xi({\Lambda})$ as follows:
\begin{equation}\label{eq:pdet}
    \mathcal{\xi}(\mathbf{\Lambda}) = \int p(d|\boldsymbol{\theta}) p(\boldsymbol{\theta}|\mathbf{\Lambda}) p_{\rm det}(d) \mathrm{d}d \mathrm{d}\boldsymbol{\theta},
\end{equation}
where $p(d|\boldsymbol{\theta})$ is the likelihood of the data $d$ given a \ac{GW} source with parameters $\boldsymbol{\theta}$ (the component mass and spins for the individual source), $p(\boldsymbol{\theta}|\mathbf{\Lambda})$ is the likelihood of this source given the population-model hyperparameters $\mathbf{\Lambda}$, and $p_{\rm det}(d)$ is the probability that the data $d$ produces a trigger above the detection threshold in a search.
In practice, $p_{\rm det}(d)$ is determined by the detection statistic assigned by the search.

The importance sampling approach from~\cite{Essick:2022ojx} is adopted to calculate $\xi(\mathbf{\Lambda})$ for various choices of $\mathbf{\Lambda}$.
We draw $N_{\rm inj}$ source properties from a fiducial population model described by $\mathbf{\Lambda}_0$, and inject simulated signals drawn from the fiducial population into data from \ac{O4a}.
Subsequently, the searches analyze this data to determine the injection recovery.
Since the injections are randomly sampled from $\mathbf{\Lambda}_0$, $\xi(\mathbf{\Lambda})$ can be computed through a Monte Carlo sum over the subset of found detections denoted by the set of \ac{GW} source parameters $\{ {\boldsymbol\theta} \}_{\rm found}$, which pass the chosen detection threshold as follows:
\begin{equation}
    \mathcal{\xi}(\mathbf{\Lambda}) \approx \frac{1}{N_{\rm inj}} \sum_{\boldsymbol\theta\in\{{\boldsymbol\theta}\}_{\rm found}} \frac{p(\boldsymbol{\theta}|\mathbf{\Lambda})}{p(\boldsymbol{\theta}|\mathbf{\Lambda}_0)},
    \label{eq:MC_efficiency}
\end{equation}
where $p(\boldsymbol{\theta}|\mathbf{\Lambda}_0)$ denotes the probability of the detected source drawn from the fiducial population ($\mathbf{\Lambda}_0$), while $p(\boldsymbol{\theta}|\mathbf{\Lambda})$ denotes the probability of the detected source given the population distribution of interest ($\mathbf{\Lambda}$).
Table~\ref{tab:injection_population} summarizes the fiducial population describing the source parameters for simulated signals.

Within each population, the simulated signals were injected \InjSpacing\,s apart.
\PYCBC and \GSTLAL filtered the \ac{BBH} and \ac{BNS} populations separately resulting in analyzing two streams of simulated data.
To analyze the simulated data in a single pass, \MBTA shifted the \ac{BNS} injections by \MBTAInjSpacing\,s relative to the \ac{BBH} simulations before injecting them in strain data, resulting in an effective spacing of \MBTAInjSpacing\,s.
The reduced spacing verifiably did not have a noticeable effect on the efficiency of the \MBTA search.

\begin{table*}[t]
    \tablewidth{\linewidth}
    \setlength{\tabcolsep}{30pt} 
    \renewcommand{\arraystretch}{1.4} 
    \caption{\label{tab:injection_population} Fiducial population of \ac{BBH} and \ac{BNS} sources used for determining the sensitive hypervolume $\langle VT \rangle$.}
    \begin{tabular}{ c  c  c }
        \hline \hline
        & BBH & BNS \\
        \hline
        \textsc{Mass Model}     & $\masscomponent^{\alpha}, \alpha=\InjMassPowerLawAlpha$   &  $\masscomponent^{\alpha}, \alpha=\InjMassPowerLawAlpha$ \\
                                & $\massone \in [\MinPrimaryMassBH, \MaxPrimaryMassBH]$     &  $\massone \in [\MinPrimaryMassNS, \MaxPrimaryMassNS]$ \\
                                & $\masstwo \in [\MinSecondaryMassBH, \MaxSecondaryMassBH]$ &  $\masstwo \in [\MinSecondaryMassNS, \MaxSecondaryMassNS]$ \\
        \hline
        \textsc{Spin Magnitude Model}     & \multicolumn{2}{c}{$w\sqrt{\frac{2}{\pi\sigma^2}}\exp\left(-\frac{\chi_{1,2}^2}{2\sigma^2}\right)
      +(1-w)\frac{\chi_{1,2}^{\alpha-1}\,(1\,-\,\chi^{}_{1,2})^{\,\beta-1}}{{\rm B}(\alpha,\,\beta)}$} \\
                                & \multicolumn{2}{c}{$\mu=\NormalDistMu, \sigma=\NormalDistSigma, \alpha = \BetaDistAlpha, \beta = \BetaDistBeta, w=\DistributionWeight$} \\
                                & \multicolumn{2}{c}{$\spincomponent  \in [\MinSpinMag,\MaxSpinMag]$} \\
        \hline
        \textsc{Waveform Model} & \IMRPHENOMXPHM & \IMRPHENOMXPNRTIDALVTHREE \\
        \hline
        \textsc{Equation of State} &  & \GPPVADDtwo \\
        \hline
  \end{tabular}
\tablecomments{Sources are drawn from the given distributions for component masses \masscomponent, and dimensionless spin magnitudes \spincomponent to perform the searches on augmented \ac{O4a} data. 
We assume isotropically distributed spins, i.e. the cosine of the angles between the orbital angular momentum vector and the component spin vectors are distributed uniformly. The sources are distributed uniformly in comoving volume and time up to a maximum redshift $z_{\rm max} =$ \MaxRedshiftInj. This population is denoted by $\mathbf{\Lambda_0}$ in Equation~(\ref{eq:MC_efficiency}).}
\end{table*}